\documentclass[12pt]{iopart}
\usepackage{graphicx}
\begin{document}
\title{Brane World Cosmology}
\author{Philippe Brax\dag~, Carsten van de Bruck\ddag\\ and Anne--Christine Davis$^{*}$}
\address{\dag\ Service de Physique Th\'eorique, CEA-Saclay
F-91191, Gif/Yvette cedex, France}
\address{\ddag\ Department of Applied Mathematics, Astro--Particle and Cosmology Group\\
~~University of Sheffield, Hounsfield Road, Hicks Building \\
~~Sheffield S3 7RH, United Kingdom}
\address{$^*$ Department of Applied Mathematics and Theoretical Physics,\\
~~Center for Mathematical Sciences, University of Cambridge, \\
~~Wilberforce Road, Cambridge CB3 0WA, U.K.}
\ead{brax@spht.saclay.cea.fr, C.vandeBruck@sheffield.ac.uk, acd@damtp.cam.ac.uk}
\begin{abstract}
Recent developments in the physics of  extra dimensions have
opened up new avenues to test such theories. We review cosmological
aspects of brane world scenarios such as the Randall--Sundrum
brane model and two--brane systems  with a bulk scalar field. We
start with the simplest brane world scenario leading to a
consistent cosmology: a brane embedded in an Anti--de Sitter space--time. 
We generalise this setting to the case with  a bulk scalar
field and then  to two--brane systems.

We discuss different ways  of  obtaining  a low--energy effective
theory for two--brane systems, such as the moduli space
approximation and the low--energy expansion. A comparison between
the different methods is given. Cosmological perturbations are
briefly discussed as well as early universe scenarios such as the
cyclic model and the born--again brane world model.
Finally we also present some physical consequences of brane world scenarios 
on the cosmic microwave background and the variation of constants.

\end{abstract}

\maketitle

\section{Introduction}

Whilst theories formulated in extra dimensions have been around
since the early twentieth century, recent developments have opened
up new avenues which have  enabled the cosmological consequences
of such theories to be extracted. This, together with advances in
high precision cosmological data, opens up the exciting possibility
of testing and constraining extra--dimensional theories for the first time. 
In this review we explore these developments.

In the early twentieth century Kaluza and Klein \cite{Kaluza},
following Nordstrom, attempted to unify electromagnetism and
gravity by proposing a theory in five space--time dimensions, with
the extra spatial dimension compactified on a circle. More
recently, attempts to construct a consistent theory incorporating
both quantum mechanics and gravity have been constructed in more
than four dimensions, with the extra, spatial dimensions
compactified. In particular superstring theories, which include
both quantum theory and gravity, are only consistent in ten
dimensions \cite{polchinskibook}, with the extra dimensions being
spatial. The usual four dimensional physics is retrieved by
compactifying on a compact manifold with dimensions typically of
the order of  the Planck scale. This is essentially a
generalisation of Kaluza--Klein theory.

Another approach has emerged recently, motivated by M-theory.
M-theory is an umbrella theory that, in certain limits, reduces to
the five known string theories or to supergravity. Horava and
Witten \cite{horavawitten} showed that the strong coupling limit
of the $E_8 \times E_8$ heterotic string theory at low energy is
described by eleven dimensional supergravity with the eleventh
dimension compactified on an orbifold with $Z_2$ symmetry, i.e. an
interval. The two boundaries of space--time (i.e.  the orbifold
fixed points) are 10--dimensional planes, on which gauge theories
(with $E_8$ gauge groups) are confined. Later Witten argued
that 6 of the 11 dimensions can be consistently compactified on a
Calabi--Yau threefold and that the size of the Calabi-Yau manifold
can be substantially smaller than the space between the two
boundary branes \cite{witten}, i.e. Planckian. The branes are
hypersurfaces embedded in the extra dimension. Thus, in this limit
space--time looks five--dimensional with four dimensional boundary
branes \cite{lukasstelle}. This has led to the concept of brane
world models.

More generally speaking, in brane world models 
the standard model particles are confined on
a hypersurface (called a brane) embedded in a higher dimensional
space (called the bulk). Only gravity and other exotic matter such as
the superstring dilaton field can propagate in the bulk. Our
universe may  be such a brane--like object. This idea was
originally motivated phenomenologically (see
\cite{akama}--\cite{gibbons}) and later revived by the recent
developments in string theory. Within the brane world scenario,
constraints on the size of extra dimensions become weaker because
the standard model particles propagate only in three spatial
dimensions. Newton's law of gravity, however, is sensitive to the
presence of extra--dimensions. Gravity is being tested only on
scales larger than a tenth of a millimeter and possible deviations
below that scale can be envisaged.

Arkani-Hamed, Dimopoulos and Dvali (ADD) \cite{arkanihamed1} and
\cite{arkanihamed2}, following an earlier idea by Antoniadis
\cite{antoniadis}, proposed an interesting, but simple model,
considering a flat geometry in ($4+d$)--dimensions, in which $d$
dimensions are compact with radius $R$ (toroidal topology). All
standard model particles are confined to the brane, but gravity
can explore the extra $d$ dimensions. This gives rise to a
modification of the gravitational force law as seen by an observer
on the brane. Two test masses, $m_{1}$ and $m_2$, at distances
$r$ apart will feel a gravitational potential of
\begin{equation}
V(r) \approx {m_{1}m_{2}\over M_{\rm fund}^{d+2}} {1\over
r^{d+1}},\  r \ll R
\end{equation}
and
\begin{equation}
V(r) \approx {m_{1}m_{2}\over M_{\rm fund}^{d+2}} {1\over {R^d
r}},\   r \gg R
\end{equation}
where $M_{\rm fund}$ is the fundamental Planck mass in the higher
dimensional space--time. This allows one to identify the
four--dimensional Planck mass $M_{\rm Pl}$,
 \begin{equation}
M_{\rm Pl}^2 = M_{\rm fund}^{2+d} R^d.
\end{equation}
>From above we see that on scales larger than $R$ gravity behaves
effectively as the usual four--dimensional gravity. However, on
scales less than $R$ there are deviations and gravity looks truly
$4+d$ dimensional. Since gravity is tested only down to sizes of
around a millimeter, $R$ could be as large as a fraction of a
millimeter.

The most important result of the ADD proposal is a possible
resolution to the hierarchy problem, that is the large discrepancy between
the Planck scale at $10^{19}$ GeV and the electroweak scale at
$100$ GeV. The fundamental Planck
mass could be comparable to the electroweak scale as long as the
volume of the extra dimensional space is large enough. To realise
this, their proposal requires more than one extra dimension.

However, progress was made by Randall and Sundrum, who considered
{\it non--flat}, i.e. warped bulk geometries
\cite{randallsundrum1}, \cite{randallsundrum2}. Their models were
formulated in one extra dimension with the bulk space--time being a
slice of Anti--de Sitter space--time, i.e. a space--time with a
negative cosmological constant. They proposed a two--brane model
in which the hierarchy problem can be addressed. The large hierarchy is due 
to the highly curved AdS background which implies a large gravitational
red-shift between the energy scales on the two branes. In this
scenario, the standard model particles are confined on a brane
with negative tension sitting at $y=r_c$, whereas a positive
tension brane is located at $y=0$, where $y$ is the extra spatial
dimension.
 The large hierarchy is generated by the appropriate inter-brane
distance, i.e. the  radion. It can be shown that the Planck mass
$M_{\rm Pl}$ measured on the negative tension brane is given by
($k = \sqrt{-\Lambda_5\kappa_5^2/6}$),
\begin{equation}
M_{\rm Pl}^2 \approx e^{2kr_c}M_5^3/k,
\end{equation}
where $M_5$ is the five--dimensional Planck mass and $\Lambda_5$
the (negative) cosmological constant in the bulk. Thus, we see
that, if $M_5$ is not very far from the electroweak scale $M_W
\approx $TeV, we need $kr_c \approx 50$, in order to generate a
large Planck mass on our brane. Hence, by tuning the radius $r_c$
of the extra dimension to a reasonable value, one can obtain a
very large hierarchy between the weak and the Planck scale. Of
course, a complete realisation of this mechanism requires an
explanation for such a value of  the radion. In other words, the
radion needs to be stabilised at a certain value. The
stabilisation mechanism is not thoroughly understood, though
models with a bulk scalar field have been proposed and have the
required properties \cite{goldbergerwise}.

This two brane model was unrealistic because standard model
particles are confined to the negative tension brane. However,
Randall and Sundrum  realised that due to the curvature of the bulk space time,
Newton's law of gravity could be obtained on a positive tension
brane embedded in an infinite extra dimension with warped
geometry. Small corrections to Newton's law are generated and
constrain the possible scales in the model to be smaller than a
millimetre. This model does not solve the hierarchy problem, but
has interesting cosmological implications. 

A spectacular consequence of brane cosmology, which we discuss in
detail in a later section, is the possible modification of the
Friedmann equation at very high energy \cite{binetruy1}. This
effect was first recognised \cite{boundaryinflation} in the
context of inflationary solutions. As we will see, Friedmann's
equation has, for the Randall--Sundrum model, the form
(\cite{grojean} and \cite{csaki})
\begin{equation}
H^2=\frac{\kappa_5^4}{36}\rho^2 + \frac{8\pi G_N}{3}\rho +
\Lambda,
\end{equation}
relating the expansion rate of the brane $H$ to the (brane) matter
density $\rho$ and the (effective) cosmological constant
$\Lambda$. The cosmological constant can be tuned to zero by an
appropriate choice of the brane tension and bulk cosmological
constant, as in the Randall-Sundrum case. Notice that at high
energies, for which
\begin{equation}
\rho \gg \frac{96 \pi G_N}{\kappa_5^4},
\end{equation}
where $\kappa_5^2$ is the five dimensional gravitational constant,
the Hubble rate becomes
\begin{equation}
H \propto \rho,
\end{equation}
while in ordinary cosmology $H\propto \sqrt \rho$. The latter case
is retrieved at low energy, i.e.
\begin{equation}
\rho \ll \frac{96 \pi G_N}{\kappa_5^4},
\end{equation}
Of course modifications to the Hubble rate can only be significant
before nucleosynthesis, though they may have drastic consequences on
early universe phenomena such as inflation.

A natural extension to the Randall--Sundrum model is to consider a
bulk scalar field. This occurs automatically in supersymmetric
extensions and makes more contact to string theory, which has the
dilaton field. The addition of a bulk scalar field allows other
fundamental problems to be addressed, for instance that of the
cosmological constant. Attemps have been made to invoke an extra--dimensional 
origin for  the apparent (almost) vanishing of the
cosmological constant. The self--tuning idea \cite{selftuning}
advocates that the energy density on our brane does not lead to a
large curvature of our universe. On the contrary, the extra
dimension becomes highly curved, preserving a flat Minkowski brane
with apparently vanishing cosmological constant. Unfortunately, the
simplest realisation of this mechanism with a bulk scalar field
fails due to the presence of a naked singularity in the bulk. This
singularity can be shielded by a second brane whose tension has to
be fine--tuned with the original brane tension \cite{nilles}. In a
sense, the fine tuning problem of the cosmological constant
reappears through the extra dimensional back-door. However, in 
supersymmetric extensions of the Randall--Sundrum model 
the bulk scalar field can give rise to a late--time acceleration of
the universe once supersymmetry is broken on the physical brane
world. The naked singularity mentioned above is automatically
beyond the second brane. There is still a fine--tuning problem
here, namely that of supersymmetry breaking. We discuss this in detail in
later sections.

One might hope that brane world models could explain the origin of
the universe itself. Attempts have been made in this direction by
considering brane collisions. In this scenario the universe
undergoes a series of big bangs and big crunches as the branes go
through cycles of moving apart and colliding. This is an exciting
development which we briefly discuss in the last section.

In this review we study the physics of one brane systems both
with an empty bulk and with a bulk scalar field. Birkhoff's theorem is
discussed and examined in the case of a bulk scalar field.  We then turn 
our attention to the case of two brane systems. The low--energy behaviour of 
two brane systems is emphasised. Throughout this 
review we discuss the cosmological implications and predictions 
of brane worlds. In particular, the CMB predictions of simple brane 
models are reviewed. Finally, we
conclude by considering brane collisions both in the cyclic and
born--again brane world scenarios. The review is an extension 
of an earlier review \cite{ourreview}, with recent developments and 
new material included. There 
are other reviews on brane worlds and their cosmological implications 
(see e.g. \cite{luki} - \cite{quevedo}), each with a different emphasis. 

\section{A brane in an anti--de Sitter bulk space--time}
We begin with a discussion of the simplest non--trivial (i.e.
different from Minkowski space--time) bulk, the Anti--de Sitter
(AdS) space--time. The AdS space--time is a maximally symmetric
solution to Einstein's equation with negative cosmological
constant.

The first well motivated brane world model with an Anti--de Sitter
bulk space--time was suggested by Randall and Sundrum. Initially, they
proposed a two--brane scenario in five dimensions with a highly
curved bulk geometry as an explanation for the large hierarchy
between the  Planck scale and the electroweak energy scale
\cite{randallsundrum1}. In their scenario the standard model
particles live on a brane with (constant) negative tension,
whereas the bulk is a slice of  Anti--de Sitter (AdS) space--time.
In the bulk there is another brane with positive tension. This is
the so--called Randall--Sundrum I (RSI) model, in order to
distinguish this setup from a one brane model, which they proposed
immediately afterwards. Analysing the solution of Einstein's
equation on the  positive tension brane and sending the negative
tension brane to infinity, an observer confined to the positive
tension brane recovers Newton's law if the curvature scale of the
AdS space  is smaller than a millimeter \cite{randallsundrum2}. Note that
the higher--dimensional space is {\it non--compact}, which must be
contrasted with the Kaluza--Klein mechanism, where all
extra--dimensional degrees of freedom are compact. This one--brane
model, the so--called Randall--Sundrum II (RSII) model, will be
our main focus in this section.

Before discussing  the cosmological consequences of the second
Randall--Sundrum model, let us first look at the {\it static}
solution, introducing some valuable techniques. The (total)
action consists of the Einstein-Hilbert action and the brane
action, which have the form
\begin{eqnarray}
S_{\rm EH} &=& -\int d^5x \sqrt{-g^{(5)}}
\left( \frac{R}{2\kappa_5^2} + \Lambda_{5} \right), \\
S_{\rm brane} &=& \int d^4x \sqrt{-g^{(4)}}\left( -\sigma \right).
\end{eqnarray}
Here, $\Lambda_5$ is the bulk cosmological constant
and $\sigma$ is the (constant) brane tension. $\kappa_5$
is the five--dimensional gravitational coupling constant. The
brane is located at $y=0$ and we assume a $Z_2$ symmetry, i.e.
we identify $y$ with $-y$. The ansatz for the metric is
\begin{equation}
ds^2 = e^{-2 K(y)}\eta_{\mu\nu} dx^{\mu}dx^{\nu} + dy^2.
\end{equation}
Einstein's equations, derived from the action above, give two
independent equations:
\begin{eqnarray}
6K'^2 &=& - \kappa_5^2 \Lambda_{5} \nonumber \\
3K'' &=& \kappa_5^2 \sigma \delta(y) \nonumber.
\end{eqnarray}
The first equation can be easily solved:
\begin{equation}\label{K}
K = K(y) = \sqrt{-\frac{\kappa_5^2}{6}\Lambda_5}\ y \equiv ky,
\end{equation}
which tells us that $\Lambda_5$ must be negative. If we integrate
the second equation from $-\epsilon$ to $+\epsilon$, take the limit
$\epsilon \rightarrow 0$ and make use of the $Z_2$ symmetry, we get
\begin{equation}
6K'\vert_0 = \kappa_5^2 \sigma
\end{equation}
Together with eq. (\ref{K}) this tells us that
\begin{equation}\label{finetuning}
\Lambda_5 = -\frac{\kappa_5^2}{6}\sigma^2
\end{equation}
Thus, there must be a fine tuning between the brane
tension and the bulk cosmological constant  for
static solutions to exist.

Randall and Sundrum have shown that  there is a {\it continuum} of
Kaluza--Klein modes for the gravitational field, which is
different from the case of a periodic  extra dimension where a
discrete spectrum is predicted. The continuum of Kaluza--Klein
modes lead to a correction to the force between two static masses
on the brane. To be more specific, it was shown that the potential
energy between two point masses confined on the brane is given by
\begin{equation}
V(r) = \frac{G_N m_1 m_2}{r}\left(1 + \frac{l^2}{r^2} + O(r^{-3})\right).
\end{equation}
As gravitational experiments show no deviation from Newton's law of
gravity on length scales larger than a millimeter \cite{submillimeter},
$l$ has to be smaller than that length scale.

A remark is in order: we assumed a $Z_2$ symmetry, i.e. we are
considering the orbifold $S_1/Z_2$. Most  brane world scenarios
assume this symmetry. The physical motivation for considering this
orbifold compactification comes from heterotic M--theory (as
mentioned in the introduction). Of course, assuming the
$Z_2$ symmetry leads to a significant simplification of the
equations and therefore simplicity is a good motivation, too.

In the following we derive Einstein's equation for an observer on the brane.

\subsection{Einstein's equations on the brane}
There are two ways of deriving the cosmological equations and we
will describe both of them below. The first one is rather simple
and makes use of the bulk equations only. The second method uses
the  geometrical relationship between four--dimensional and
five--dimensional quantities. We begin with the simpler method.

\subsubsection{ Friedmann's equation from
five--dimensional Einstein equations}
In the following subsection we will set $\kappa_5 \equiv 1$ as this
simplifies the equations a lot. We write the bulk metric as follows:
\begin{equation}\label{metricbraneframe}
ds^2 = a^2 b^2(dt^2 - dy^2) - a^2 \delta_{ij}dx^i dx^j.
\end{equation}
This metric is consistent with homogeneity and isotropy
on the brane located at $y=0$. The functions $a$ and $b$ are
functions of $t$ and $y$ only. Furthermore, we have assumed
flat spatial sections, but it is straightforward to include a
spatial curvature. For this metric, Einstein equations
in the bulk read:
\begin{eqnarray}
a^2b^2{G^0}_0 &\equiv& 3\left[2\frac{\dot{a}^2}{a^2}+\frac{\dot{a}\dot{b}}{ab}
                   -\frac{a''}{a}+\frac{a' b'}{ab}+kb^2\right]
                = a^2b^2\left[\rho_B +\rho^{ }
                \bar\delta(y-y_b)\right]\label{G00}\\
a^2b^2{G^5}_5 &\equiv&3\left[\frac{\ddot{a}}{a}-\frac{\dot{a}\dot{b}}{ab}
                 -2\frac{{a'}^2}{a^2}-\frac{a' b'}{ab}+kb^2\right]
                 =-a^2b^2T^5_5 \label{G55}\\
a^2b^2{G^0}_5&\equiv&3\left[ -\frac{\dot{a}'}{a}+2\frac{\dot{a}a'}{a^2}
                 +\frac{\dot{a}b'}{ab}+\frac{a'\dot{b}}{ab}\right]
                 = -a^2b^2 T^0_5  \label{G05}\\
a^2b^2{G^i}_j&\equiv&\left[ 3\frac{\ddot{a}}{a}+\frac{\ddot{b}}{b}-
                 \frac{\dot{b}^2}{b^2}-3\frac{a''}{a}-\frac{b''}{b}
                 +\frac{{b'}^2}{b^2}+kb^2\right]
                 {\delta^i}_j \nonumber \\
                 &=&-a^2b^2\left[p_B+p^{ }
                 \bar{\delta}(y-y_b)\right]{\delta^i}_j ,\label{Gij}
\end{eqnarray}
where the bulk energy--momentum tensor $T^a_b$ has been kept
general here. For the Randall--Sundrum model we will now take
$\rho_B = - p_B = \Lambda_5$ and $T^0_5=0$. Later  we will make
use of these equations to derive Friedmann's equation with a bulk
scalar field. In the equations above, a dot represents the
derivative with respect to $t$ and a prime a derivative with
respect to $y$.

If one integrates the 00--component over $y$ from $-\epsilon$ to
$\epsilon$ and use make use of the $Z_2$ symmetry (which implies
here that $a(y)=a(-y)$, $b(y)=b(-y)$, $a'(y)=-a'(-y)$ and $b'(y)=-b'(-y)$),
then, in the limit $\epsilon \rightarrow 0$, one gets
\begin{equation}\label{junc1}
\frac{a'}{a}\left|_{y=0} \right. = \frac{1}{6}ab \rho.
\end{equation}
Similarly, integrating the $ij$--component and using the last
equation gives
\begin{equation}\label{junc2}
\frac{b'}{b}\left|_{y=0} \right. = -\frac{1}{2}ab (\rho+p).
\end{equation}
These two conditions are called the junction conditions and
play an important role in cosmology, describing
how the brane with some given energy--momentum tensor $T_{\mu\nu}$ can be
embedded in a higher--dimensional space--time. The other
components of the  Einstein  equations should be compatible with these
conditions. It is not difficult to show that the restriction of
the $05$ component to $y=0$ leads to
\begin{equation}\label{energyRS}
\dot\rho + 3\frac{\dot a}{a}\left(\rho + p \right) = 0,
\end{equation}
where we have made use of the junction conditions (\ref{junc1}) and
(\ref{junc2}). This is nothing but matter conservation on the
brane, which must hold because we have assumed that matter is confined
on the brane.

Restricting the 55--component of the bulk equations to the brane and
using again the junction conditions gives
\begin{equation}
\frac{\ddot{a}}{a}-\frac{\dot{a}\dot{b}}{ab}+kb^2
  = -{a^2b^2\over 3}\left[\frac{1}{12}\rho^{ }\left(\rho{ }+3p^{ }\right)
    +q_B\right]\; .
\end{equation}
Changing to cosmic time $d\tau = ab dt$, writing
$a = \exp(\alpha(t))$ and using the energy conservation gives (\cite{flanagan},
\cite{vandebruck1})
\begin{equation}
\frac{d(H^2 e^{4\alpha})}{d\alpha} =
\frac{2}{3}\Lambda_5 e^{4\alpha}
+ \frac{d}{d\alpha}\left( e^{4\alpha} \frac{\rho^2}{36}\right).
\end{equation}
In this equation $aH=da/d\tau$. This equation can easily be
integrated to give
\begin{equation}
H^2 = \frac{\rho^2}{36} + \frac{\Lambda_5}{6} + \frac{\mu}{a^4}.
\end{equation}
Let us split the total energy density and pressure into
parts coming from matter and brane tension, i.e. we write
$\rho = \rho_M + \sigma$ and $p = p_M - \sigma$.
Then we obtain Friedmann's equation
\begin{equation}
H^2 = \frac{8\pi G}{3}\rho_M\left[1+\frac{\rho_M}{2\sigma}\right] +
\frac{\Lambda_4}{3} + \frac{\mu}{a^4},
\end{equation}
where we have made the identification
\begin{eqnarray}
\frac{8\pi G}{3} &=& \frac{\sigma}{18} \\
\frac{\Lambda_4}{3} &=& \frac{\sigma^2}{36} + \frac{\Lambda_5}{6}.
\end{eqnarray}
Imposing the fine--tuning relation (\ref{finetuning}) of the
static Randall--Sundrum solution in the last equation, we see that
$\Lambda_4 = 0$. If there is a small mismatch between the brane tension and
the five--dimensional cosmological constant, an effective
four--dimensional cosmological constant is generated. Hence, the
Randall--Sundrum setup does not provide a solution to the cosmological
constant problem, as one has to impose a relation between brane tension and
the cosmological constant in the bulk, i.e. equation (\ref{finetuning}).

Another result is that the four--dimensional Newton constant
is directly related to the brane tension in this model.

The constant $\mu$ appears in the derivation above as an integration constant.
The term including $\mu$ is sometimes called the {\it dark radiation} term
(see e.g. \cite{darkradiation1}-\cite{darkradiation3}).
The parameter $\mu$ can be obtained from a full analysis of the
bulk equations \cite{shirubulk}-\cite{binetruy2} (we will discuss this
in section 4). An extended version of Birkhoff's theorem tells us
that, if the bulk space--time is AdS, this constant is zero
\cite{charmousis}. If
the bulk is AdS--Schwarzschild instead, $\mu$ is non--zero but a
measure of the mass of the bulk black hole. In the following we will assume
that $\mu=0$ and $\Lambda_4=0$.

The most significant change in Friedmann's equation compared to the
usual four--dimensional form is the appearance of a term proportional
to $\rho^2$. This term implies that if the matter energy density is much
larger than the brane tension, i.e. $\rho_M \gg \sigma$, the expansion
rate $H$ is proportional to $\rho_M$, instead of $\sqrt{\rho_M}$. The expansion
rate is, in this regime, larger in this brane world scenario. Only in the
limit where the brane tension is much larger than the matter energy density is
the usual behaviour $H \propto \sqrt{\rho_M}$ recovered. This is the most
important change in brane world scenarios. As we will see later, it is quite
generic and not restricted to the Randall--Sundrum brane world model.

At the time of nucleosythesis the brane world corrections in
Friedmann's equation must be negligible, otherwise the expansion
rate is modified and the results for the abundances of the light
elements are unacceptably changed. This implies that $\sigma \ge
(1 {\rm MeV})^4$. Note, however, that a much stronger constraint
arises from current tests for deviation from Newton's law
\cite{maartensw} (assuming the Randall--Sundrum fine--tuning
relation (\ref{finetuning})): $\kappa_5^{-3} > 10^5$ TeV and
$\sigma \ge (100 {\rm GeV})^4$. Similarily, cosmology constrains
the amount of dark radiation. It has been shown that the energy
density in dark radiation can be at most be 10 percent of the
energy density in photons \cite{darkradiation3}.

Finally we can derive Raychaudhuri's equation from Friedmann's equation and the
energy--conservation equation:
\begin{equation}
\frac{dH}{d\tau} = - 4\pi G (\rho_M + p_M)\left[ 1+ \frac{\rho_M}{\sigma}
\right].
\end{equation}

\subsubsection{Another derivation of Einstein's equation}
In the following, we present a more geometrical and, hence, more
powerful derivation of Einstein's equation on the brane
\cite{shiromizu}.

For this, consider an arbitrary (3+1) dimensional
hypersurface ${\cal M}$ with unit normal vector $n_a$ embedded
in a five--dimensional space--time. The induced metric and the
extrinsic curvature of the hypersurface are
defined as
\begin{eqnarray}
h^a_{~b} &=& \delta^a_{~b} - n^a n_b, \\
K_{ab} &=& h_a^{~c}h_b^{~d} \nabla_c n_d.
\end{eqnarray}
For the derivation we need three equations; two of them relate
four--dimensional quantities constructed from $h_{ab}$ to
full five--dimensional quantities constructed from $g_{ab}$.
We just state these equations here and refer to \cite{wald} for
their derivation.
The first equation is the Gauss equation, which reads
\begin{equation}
R^{(4)}_{abcd} = h_{a}^{~j}h_{b}^{~k} h_{c}^{~l}h_{d}^{~m}
R_{jklm} - 2K_{a[c}K_{d]b}.
\end{equation}
This equation relates the four--dimensional curvature tensor
$R^{(4)}_{abcd}$, constructed from $h_{ab}$, to the five--dimensional
one and $K_{ab}$. The next equation is the Codazzi equation, which
relates $K_{ab}$, $n_a$ and the five--dimensional Ricci tensor:
\begin{equation}
\nabla^{(4)}_b K^{b}_{~a} - \nabla^{(4)}_a K
= n^c h^{b}_{~a}R_{bc}.
\end{equation}
One decomposes the five--dimensional curvature
tensor $R_{abcd}$ into the Weyl tensor $C_{abcd}$ and the
Ricci tensor:
\begin{equation}
R_{abcd} = \frac{2}{3} \left(g_{a[c}R_{d]b}
- g_{b[c}R_{d]a} \right) - \frac{1}{6} R g_{a[c}g_{b]d}
+ C_{abcd}.
\end{equation}
If one substitutes  the last equation into the Gauss equation and
constructs the four--dimensional Einstein tensor, one obtains
\begin{eqnarray}\label{eintensor}
G^{(4)}_{ab} &=& \frac{2}{3}\left( G_{cd} h^{c}_{~a}h^{d}_{~b}
 + \left(G_{cd}n^c n^d - \frac{1}{4}G \right)h_{ab} \right)
\nonumber \\
&+& K K_{ab} - K_a^{~c}K_{bc} - \frac{1}{2}\left(K^2
- K^{cd}K_{cd}\right)h_{ab} - E_{ab}~,
\end{eqnarray}
where
\begin{equation}
E_{ab} = C_{acbd}n^c n^d.
\end{equation}
We would like to emphasise that this equation holds for any
hypersurface. If one considers a hypersurface with energy
momentum tensor $T_{ab}$, then there exists a relationship
between $K_{ab}$ and $T_{ab}$ ($T$ is the trace of $T_{ab}$) \cite{israel}:
\begin{equation}
\left[ K_{ab} \right] = -\kappa_5^2 \left(T_{ab}
- \frac{1}{3} h_{ab}T \right),
\end{equation}
where $[...]$ denotes the {\it jump}:
\begin{equation}
[f](y) = {\rm lim}_{\epsilon \rightarrow 0}
\left( f(y+\epsilon) - f(y-\epsilon)\right).
\end{equation}
These equations are called junction conditions and are
equivalent, in the cosmological context, to the junction
conditions (\ref{junc1}) and (\ref{junc2}). Splitting
$T_{ab} = \tau_{ab} - \sigma h_{ab}$ and inserting the junction
condition into equation (\ref{eintensor}), we obtain
Einstein's equation on the brane:
\begin{equation}\label{einsteinbrane}
G^{(4)}_{ab} = 8\pi G \tau_{ab} - \Lambda_4 h_{ab}
+ \kappa_5^4 \pi_{ab} - E_{ab}.
\end{equation}
The tensor $\pi_{ab}$ is defined as
\begin{equation}
\pi_{ab} = \frac{1}{12} \tau \tau_{ab} - \frac{1}{4}
\tau_{ac}\tau_{b}^{~c} + \frac{1}{8} h_{ab}\tau_{cd}\tau^{cd}
- \frac{1}{24} \tau^2 h_{ab},
\end{equation}
whereas
\begin{eqnarray}
8\pi G &=& \frac{\kappa_5^4}{6}\sigma \\
\Lambda_4 &=& \frac{\kappa_5^2}{2}\left(\Lambda_5 +
\frac{\kappa_5^2}{6}\sigma^2\right).
\end{eqnarray}
Note that in the Randall--Sundrum case we have $\Lambda_4 = 0$
due to the fine--tuning between the brane tension and the bulk cosmological
constant. Moreover  $E_{ab} = 0$ as  the Weyl--tensor vanishes for
an AdS space--time. In general, energy conservation and the
Bianchi identities imply that
\begin{equation}\label{weylfluid}
\kappa_5^4 \nabla^{a} \pi_{ab} = \nabla^{a} E_{ab}
\end{equation}
on the brane.

Clearly, this method is very powerful, as it does not assume
homogeneity and isotropy nor does it assume the bulk to be AdS.
In the case of an AdS bulk and a Friedmann--Robertson--Walker brane,
the previous equations  reduce to the Friedmann equation and
Raychaudhuri equation derived earlier.

To conclude this section, we have given two derivations of Einstein's
equation as perceived by an observer confined on a brane embedded
in an Anti--de Sitter bulk. In the derivation the emphasis was
given to the brane observer. This point of view, however, neglects
the role of the bulk. In particular it does not say anything about
the magnitude of $E_{ab}$ once the brane becomes
inhomogeneous. In fact, the set of equations on the brane {\it are
not closed in general} \cite{roy}. This is in particular a problem
when discussing the evolution of cosmological perturbations: the
evolution of the bulk gravitational field can not be solved using
Einstein's equation on the brane alone, needing instead a full
five--dimensional analysis. We will come back to this point in a later section.

\subsection{Slow--roll inflation on the brane}
We have seen that the Friedmann equation on a brane is
drastically modified at high energy where the $\rho^2$ term dominates.
As a result the early universe cosmology in brane world models tends to be
different from that of standard four--dimensional cosmology. Thus, it seems
natural to look for brane effects on early universe phenomena
such as inflation (see in particular \cite{wandsinflation} and
\cite{copeland}) and on phase--transitions \cite{anne}.

Let us therefore consider a scalar field confined on the brane.
The energy density and the pressure of this field are given by
\begin{eqnarray}
\rho_{\phi} &=& \frac{1}{2}\phi_{,\mu}\phi^{,\mu} + V(\phi),\\
p_{\phi} &=& \frac{1}{2}\phi_{,\mu}\phi^{,\mu} - V(\phi),
\end{eqnarray}
where $V(\phi)$ is the potential energy of the scalar field. The full
evolution of the scalar field is described by the (modified) Friedmann
equation, the Klein--Gordon equation and the Raychaudhuri equation.

\subsubsection{Single field slow--roll inflation}

In the following we will discuss the case of slow--roll inflation. Therefore,
the evolution of the fields is governed by (from now on a dot stands
for the derivative with respect to cosmic time)
\begin{eqnarray}
3H\dot\phi &\approx& - \frac{\partial V}{\partial \phi} \\
H^2 &\approx& \frac{8\pi G}{3} V(\phi)\left(1 + \frac{V(\phi)}{2\sigma}\right).
\end{eqnarray}
It is easy to show that these equations imply that the
slow--roll parameters are given by
\begin{eqnarray}
\epsilon &\equiv& -\frac{\dot H}{H^2}
= \frac{1}{16\pi G}\left(\frac{V'}{V}\right)^2\left[
\frac{4\sigma(\sigma + V)}{(2\sigma + V)^2} \right] \label{epsilon}\\
\eta &\equiv& \frac{V''}{3H^2}=\frac{1}{8\pi G}\left(\frac{V''}{V}\right)\left[
\frac{2\sigma}{2\sigma + V}\right]\label{eta}.
\end{eqnarray}
The modifications to General Relativity are contained in the square
brackets of these expressions. They imply that {\it for a given
potential and given initial conditions for the scalar field the
slow--roll parameters are suppressed compared to the predictions made
in General Relativity.} In other words, {\it brane world effects
ease slow--roll inflation} \cite{wandsinflation}. In the limit
$\sigma \ll V$ the parameters are heavily suppressed. It implies
that steeper potentials can be used to drive slow--roll inflation
\cite{copeland}. What does this mean for cosmological perturbations
generated during inflation? In order to study cosmological perturbations,
one needs to have information about the projected Weyl tensor, which
encodes the dynamics of the bulk space--time. As already said, one needs
a full five--dimensional analysis to study the evolution of $\delta E_{\mu\nu}$.
However, one can get some information by neglecting the backreaction of the bulk, i.e.
by neglecting the contribution of $E_{\mu\nu}$.
Let us then consider scalar perturbations first. The perturbed line
element on the brane has the usual four--dimensional form
\begin{equation}
ds^2 = -(1 + 2A)dt^2 + 2\partial_{i}B dtdx^i
+ ((1-2\psi)\delta_{ij} + D_{ij}E)dx^i dx^j,
\end{equation}
where the functions $A$, $B$, $E$ and $\psi$ depend on $t$ and
$x^i$. We will make use of the gauge invariant quantity \cite{bardeen}
\begin{equation}\label{zeta}
\zeta = \psi + H \frac{\delta \rho}{\dot \rho}.
\end{equation}
In General Relativity, the evolution equation for $\zeta$
can be obtained from the energy conservation equation \cite{malik},
which reads on large scales
\begin{equation}\label{zetaevolution}
\dot\zeta = -\frac{H}{\rho+p}\delta p_{\rm nad}.
\end{equation}
In this equation $\delta p_{\rm nad} = \delta p_{tot} - c_s^2\delta \rho$
is the non-adiabatic pressure perturbation. Thus, in the absence of
non--adiabatic pressure perturbations the quantity $\zeta$ is conserved
on superhorizon scales. The energy conservation equation holds for the
Randall--Sundrum model as well. Therefore, eq. (\ref{zetaevolution})
is still valid for the brane world model we consider. For inflation
driven by a single scalar field, $\delta p_{\rm nad}$ vanishes and therefore
$\zeta$ is constant on superhorizon scales during inflation.
Its amplitude is given in terms of the fluctuations in the scalar
field on spatially flat hypersurfaces:
\begin{equation}\label{zetafield}
\zeta = \frac{H\delta \phi}{\dot\phi}
\end{equation}
The quantum fluctuation in the (slow--rolling) scalar field obeys
$\langle (\delta\phi)^2 \rangle \approx (H/2\pi)^2$, as  the
Klein--Gordon equation is not modified in the brane world model we
consider. The amplitude of scalar perturbations is \cite{liddlebook}
$A_S^2 = 4 \langle \zeta^2 \rangle/25$. Using the slow--roll equations
and  eq. (\ref{zetafield}) one obtains \cite{wandsinflation}
\begin{equation}\label{scalaramplitude}
A_S^2 \approx \left( \frac{512 \pi}{75 M_{\rm Pl}^6} \right)
\frac{V^3}{V'^2} \left[\frac{2\sigma + V}{2\sigma}\right]^3
|_{k=aH}
\end{equation}
Again, the corrections are contained in the terms in the square
brackets. For a given potential the amplitude of scalar perturbations
is {\it enhanced} compared to the prediction of General Relativity.

The arguments presented so far suggest that, at least for scalar
perturbations, perturbations in the bulk space--time are not important
during inflation. This, however, might not be true for tensor
perturbations, as gravitational waves can propagate into the bulk.
For tensor perturbations, a wave equation for a single variable
can be derived \cite{langloisperturbations}.
The wave equation can be separated into a four--dimensional and a
five--dimensonal part, so that the solution has the form
$h_{ij} = A(y) h(x^\mu) e_{ij}$, where $e_{ij}$ is a (constant)
polarisation tensor. One finds that the amplitude for the zero mode
tensor perturbation is given by \cite{langloisperturbations}
\begin{equation}\label{tensoramplitude}
A_T^2 = \frac{4}{25 \pi M_{\rm Pl}^4} H^2 F^2(H/\mu)|_{k=aH},
\end{equation}
with
\begin{equation}
F(x) = \left[ \sqrt{1+x^2} - x^2 \sinh^{-1}\left(\frac{1}{x}\right)
\right]^{-1/2},
\end{equation}
where we have defined
\begin{equation}
\frac{H}{\mu} = \left( \frac{3}{4\pi\sigma} \right)^{1/2} H M_{\rm Pl}.
\end{equation}

It can be shown that modes with $m>3H/2$ are generated but they
decay during inflation. Thus, in this scenario one expects only the massless
modes to survive until the end of inflation \cite{langloisperturbations},
\cite{rubakovperturbations}. From eqns. (\ref{tensoramplitude}) and
(\ref{scalaramplitude}) one sees that the amplitudes of scalar and
tensor perturbations are enhanced at high energies, though
scalar perturbations are more enhanced than tensors. Thus, the
relative contribution of tensor perturbations will be suppressed
if inflation is driven at high energies.

Interestingly, to first order in the slow--roll parameters,
the consistency relation between the amplitudes of scalar and
tensor spectra and the spectral index of the tensor power
spectrum
\begin{equation}
\frac{A_T^2}{A_S^2} = \frac{n_T}{2}
\end{equation}
found in General Relativity holds also for inflation driven on the
brane in AdS \cite{lidsey}. However, it was shown in \cite{calgagni1}-\cite{seery} 
that this degeneracy is broken at second order in the slow--roll parameters.
Nevertheless, it will be difficult to distinguish observationally between
models of inflation in General Relativity and the models of brane world inflation
considered here. For a recent discussion on constraints on brane world inflation, see 
\cite{shinji}.

\subsubsection{Two field slow--roll inflation}
Let us consider now the case for two slowly rolling scalar fields, $\phi$ and
$\chi$, driving a period of inflation in the brane world model we consider \cite{ashcroft1}. 
To ensure the slow roll behaviour of the fields, the slow--roll parameters
\begin{eqnarray}
\epsilon_{\phi} &\equiv& \frac{1}{16\pi G}\left(\frac{V_{\phi}}{V}\right)^2\left[
\frac{4\lambda(\lambda + V)}{(2\lambda + V)^2} \right], \nonumber \\
\epsilon_{\chi} &\equiv& \frac{1}{16\pi G}\left(\frac{V_{\chi}}{V}\right)^2\left[
\frac{4\lambda(\lambda + V)}{(2\lambda + V)^2} \right], \nonumber \\
\eta_{\phi\phi} &\equiv& \frac{1}{8\pi G}\left(\frac{V_{\phi\phi}}{V}\right)\left[
\frac{2\lambda}{2\lambda + V}\right] ,\\
\eta_{\phi\chi} &\equiv& \frac{1}{8\pi G}\left(\frac{V_{\phi\chi}}{V}\right)\left[
\frac{2\lambda}{2\lambda + V}\right], \nonumber \\
\eta_{\chi\chi} &\equiv& \frac{1}{8\pi G}\left(\frac{V_{\chi\chi}}{V}\right)\left[
\frac{2\lambda}{2\lambda + V}\right], \nonumber
\end{eqnarray}
should be small. However, instead of working with the fields $\phi$ and $\chi$ 
directly, it is more convenient to consider the so--called adiabatic $\sigma$ 
and entropy field $s$. Fluctuations in $s$ are related to entropy perturbations, 
whereas perturbations in $\sigma$ describe purely adiabatic perturbations 
\cite{gordon}. The field $\sigma$ and $s$ are defined through a field rotation:
\begin{eqnarray}
\delta \sigma &=& (\cos \theta) \delta\phi   +   (\sin \theta) \delta\chi \\
\delta s &=& -  (\sin \theta) \delta\phi + (\cos \theta) \delta\chi,
\end{eqnarray}
with
\begin{equation}
\cos \theta = \frac{\dot\phi}{\sqrt{\dot\phi^2 + \dot\chi^2}}, \hspace{1cm}
\sin \theta = \frac{\dot\chi}{\sqrt{\dot\phi^2 + \dot\chi^2}}.
\end{equation}

The slow--roll parameters for $\sigma$ and $s$ are given by
\begin{equation}\label{epsilongr}
\epsilon = \frac{1}{16\pi G}\left(\frac{V_{\sigma}}{V}\right)^2
\approx \epsilon_\phi + \epsilon_\chi.
\end{equation}
and
\begin{eqnarray}
\eta_{\sigma\sigma} &=& \eta_{\phi\phi}\cos^2\theta +
2\eta_{\phi\chi}\cos\theta\sin\theta +
\eta_{\chi\chi} \sin^2 \theta, \nonumber \\
\eta_{s s} &=& \eta_{\phi\phi}\sin^2\theta -
2\eta_{\phi\chi}\cos\theta\sin\theta +
\eta_{\chi\chi} \cos^2 \theta,\label{etagr} \\
\eta_{\sigma s} &=&
(\eta_{\chi\chi}-\eta_{\phi\phi})\sin\theta\cos\theta
+ \eta_{\phi\chi}(\cos^2 \theta - \sin^2 \theta). \nonumber
\end{eqnarray}

It can be shown that the total curvature perturbation $\zeta$ and
the entropy perturbation
\begin{equation}
{\cal S} = \frac{H \delta s}{\dot \sigma}
\end{equation}
obey the equations \cite{ashcroft1}
\begin{eqnarray}
\dot{{\zeta}} &\simeq& - 2 H \eta_{\sigma s}{\cal S} \\
\dot{{\cal S}} &\simeq& \left[ -2\epsilon -\eta_{ss}
+ \eta_{\sigma\sigma} \right]H{\cal S}.
\end{eqnarray}

These are the same expressions as found in General Relativity \cite{riotto}. 
Note, however, that at very high energies, i.e. $V \gg \lambda$, we have that
the ratios $\dot{{\zeta}}/{\cal S}$ and $\dot{{\cal S}}/{\cal S}$
are suppressed by a factor $\approx \sqrt{\lambda/V}$, when compared to
General Relativity. It implies that, for given potentials and initial conditions,
entropy perturbations have less
influence on adiabatic perturbations. Even if entropy perturbations are
generated, the final correlation between the adiabatic and entropy modes
would be suppressed.

\subsection{Primordial black holes on the brane}
Let us now turn our attention to another interesting consequence of brane 
worlds:
the modification of the evaporation law of primordial black holes, which results in a higher lifetime
and lower temperature at evaporation \cite{emparan},\cite{guedens1}. It was shown that the
mass--temperature relation for a black hole
confined on the brane with radius $r_0$ much smaller than the AdS curvature length
scale is modified. Consider such a small black hole on the brane. Near the horizon,
the metric is (approximately) that of a 5D Schwarzschild black hole, which reads,
using spherical coordinates ($f = 1 - r_0^2/r^2$)
\begin{equation}
ds_5^2 = -f(r)dt^2 + f^{-1}(r)dr^2 + r^2 d\Omega_3^2.
\end{equation}
The near--horizon geometry induced on the brane is given by
\begin{equation}
ds_4^2 = -f(r)dt^2 + f^{-1}(r)dr^2 + r^2 d\Omega_2^2,
\end{equation}
which is different from the four--dimensional Schwarzschild metric. However, as in General
Relativity, it is the vicinity of the horizon which determines the thermodynamic quantities
such as entropy and temperature. For the temperature one finds
\begin{equation}
T_{\rm BH} = \sqrt{\frac{3}{32\pi}}\left(\frac{M_5^3}{M}\right)^{1/2}.
\end{equation}
Note that $T_{\rm BH} \propto 1/\sqrt{M}$ instead of $T_{\rm BH} \propto 1/M$ as in General
Relativity. Furthermore, it can be shown that Stefan's law for a D--dimensional black hole 
reads \cite{guedens1}
\begin{equation}
\frac{dM}{dt} \approx g_D \sigma_D A_{\rm eff, D} T^D, 
\end{equation}
in which $g_D$ denotes the total number of effective  bosonic and fermionic 
degrees of freedom, $\sigma_D$ is the D--dimensional Stefan--Boltzmann 
constant defined per degree of freedom ($\zeta(D)$ is the Riemann zeta 
function and $\Omega_{D-2}$ the volume of a $D-2$ sphere) 
\begin{equation}
\sigma_D = \frac{\Omega_{D-2}}{4(2\pi)^{D-1}} \Gamma(D) \zeta(D)
\end{equation}
and $A_{\rm eff, D}$ is the effective black hole area calculated from the {\it induced} 
metric. 

This implies that, for a given mass, the lifetime $t_{\rm evap} \approx M/\dot M$ 
of such a small black hole on
the brane can be much longer than black holes in General Relativity. One finds that, for a
small five--dimensional black hole,
\begin{equation}
t_{\rm evap,5D} \approx t_{\rm evap,4D} \left(\frac{l}{r_0}\right)^2,
\end{equation}
where $t_{\rm evap,4D}$ is the lifetime of a black hole of the same mass as
predicted with General Relativity. In fact, in \cite{guedens1},\cite{guedens2} it was shown that such a small
black hole can survive until the present epoch.

Once a small black hole has formed, it will accrete matter and radiation surrounding it.
It is thought that in the standard cosmology this process does not lead to a significant
enhancement of the black hole mass \cite{hawking}. However, this is no longer true in brane world
cosmology in the high energy regime, where Friedmann's equation is modified. In this regime
the radiation density obeys $\rho_{\rm rad} \propto t^{-1}$ instead of
$\rho_{\rm rad} \propto t^{-2}$. As a result, the growth in the 
mass of the black hole is $\dot M \propto t^{-1}$ (in General Relativity one
finds $\dot M \propto t^{-2}$). As
discussed in depth in \cite{guedens2}, the consequence of this is that a small black hole
can experience significant growth in the high energy regime, albeit with 
details depending on the accretion efficiency. Once the high energy regime is 
over, the accretion history is
the same as in standard cosmology. In \cite{guedens3} one finds a detailed 
discussion on how these
considerations modify some of the known constraints on primordial black holes.

For more work on black holes in brane cosmology, see \cite{bhfirst}-\cite{bhlast}. 
For reviews, see \cite{maartens} and \cite{bhreview}.

\subsection{Graviton Backreaction}
We now discuss the production of gravitons by processes on the brane \cite{graviton1},
\cite{graviton2}.
As branes carry matter fields and gravity couples to matter, one can
envisage the radiation of gravitons into the bulk. The effect of the graviton
emission is to modify the background geometry. In turn, this leads to the
expected dark radiation term on the brane. As a first step we will neglect
the back--reaction of gravitons impinging on the brane. This is much more
involved and will be briefly touched upon at the end of the subsection.

\subsubsection{Dark Radiation}

Physically, matter particles on the brane can collide and lead to
the presence of bulk gravitons. As such the brane gravitons carry
an energy--momentum tensor and therefore lead to a modification of
the bulk Einstein equations. Let us first assume that the
gravitons escape radially from the brane. The energy--momentum
tensor in the bulk is then
\begin{equation}\label{mom}
T_{ab}=\sigma k_ak_b
\end{equation}
where $k_a$ is the null impulsion of the gravitons. The general
situation when radiation is not radial will be dealt with later.
One can choose the ansatz
\begin{equation}
ds^2= -f(r,v) dv^2 +2dr dv + r^2 dx^2
\end{equation}
where
\begin{equation}
f(r,v)= k^2 r^2 -\frac{C(v)}{r^2}
\end{equation}
for the bulk metric. This is the so--called Vaidya metric. The
function $C(v)$ generalises the black hole mass as it occurs in
the static solutions obeying  Birkhoff's theorem. The Weyl
parameter $C(v)$ satisfies
\begin{equation}
\frac{dC}{dv}= \frac{2\kappa_5^2\sigma }{3} r^3 \left(\dot r - \sqrt {f+\dot r ^2}\right)^2
\end{equation}
in terms of the proper time on the brane.
The brane junction conditions are now modified and lead to ($H= \dot r/r$)
\begin{equation}
H^2=\frac{\kappa_5^2}{18}\lambda\rho +\frac{\kappa_5^2}{36} \rho^2 + \frac{C(v)}{a^4}
\end{equation}
This is the Friedmann equation where the Weyl parameter $C(v)$
plays the role of the  dark radiation term. Similarly the total
matter and pressure contents, $\rho_m= \lambda +\rho$ and $p_m=
-\lambda +p$, satisfy a non--conservation equation
\begin{equation}
\dot \rho_m +3 \frac{\dot r}{r}\left( \rho_m + p_m \right )=-2\sigma
\end{equation}
One can see that the effect of the graviton emission is to induce a loss term
in the conservation equation. The cosmology on the brane is now highly modified
by the graviton radiation term. To go further we need to evaluate $\sigma$.

Gravitons in the bulk are produced according to the tree level
process $\psi +\bar \psi \to \hbox{graviton}$ where $\psi$ and
$\bar\psi$ are particles and antiparticles confined to the brane.
As long as the temperature on the brane is much lower than
both $\kappa_5^{-2/3}$ and $\kappa_4^{-1}$, then the temperature $T$ is
always bigger than the Hubble rate. Indeed at high energy $H\sim
\kappa_5^2 T^4$ implies that $H \ll T$ and similarly at low energy
$H\sim \kappa_4 T^2$ implying that $H \ll T$ in this regime as well. We have used
Stefan's law whereby $\rho \sim T^4$ in the radiation epoch. Hence
cosmological effects can be neglected and the scattering cross
sections can be computed in Minkowski space on the brane. The
interaction Lagrangian between matter fields and gravitons is
given by
\begin{equation}
S= \kappa_5 \int dm u_m(0) \int d^4 x \tau^{\mu\nu}
h_{\mu\nu}^{(m)}
\end{equation}
where $m$ labels the continuous spectrum of Kaluza--Klein states.
The wave function of the gravitons in the extra dimension is
$u_m(y)$ corresponding to the four dimensional field
$h_{\mu\nu}^{(m)}$. Here $\tau_{\mu\nu}$ is the energy momentum of
particles on the brane. The matrix element corresponding to this
tree level interaction can be computed (averaged over spins)
\begin{equation}
\vert {\cal M}\vert = \kappa_5^2 \vert u_m(0)\vert A\frac{s^2}{8}
\end{equation}
where $s$ is the Mandelstam variable $s= (p_1+p_2)^2$
corresponding to the annihilation of particles with momenta
$p_{1,2}$. The factor $A$ is given by $A=2/3$ for scalars, $A=1$
for fermions and $A=4$ for photons.

The Boltzmann equation in curved space is
\begin{equation}
\dot \rho +3H (\rho+p) =-\int dm \int \frac{d^3 p}{(2\pi)^3 }C_m
\end{equation}
where the collision term reads
\begin{equation}
C_m= \frac{1}{2} \int
\frac{d^3p_1}{(2\pi)^3}\frac{d^3p_2}{(2\pi)^3}\vert {\cal M}\vert
^2 f_1 f_2 \delta ^{(4)} (p_m-p_1-p_2)
\end{equation}
where $f_i=1/(e^{E_i/T}\pm 1)$ depending on the statitics of the
particles. Here $p_m$ is the momentum of the gravitons. Neglecting
the contribution of the light gravitons $m \ll k$, and using $
u_m(0)= 1/\sqrt \pi$ for $m \gg k$ one finds that
\begin{equation}
\dot \rho +4H \rho = -\left[ \frac{315 \zeta ( 9/2)\zeta
(7/2)}{512 \pi^2}\right] g(T) \kappa_5^2 T^8
\end{equation}
where the effective number of species is given by
\begin{equation}
g(T)=\frac{2g_s}{3} + 4g_v + (1-2^{-7/2})(1-2^{-5/2}) g_f
\end{equation}
Here $g_s$, $g_f$ and $g_v$ are the number of relativistic
particles of spin $0, 1/2 $ and $1$.

This allows one to identify the parameter $\sigma$
\begin{equation}
\sigma= \frac{\alpha}{12} \kappa_5^2 \rho^2
\end{equation}
where Stefan's law reads $\rho= \frac{\pi^2}{30} g_* T^4$ where
$g_*= g_s + g_v +7/8 g_f$ and we have defined the parameter
\begin{equation}
\alpha= \frac{212625}{64 \pi^7} \zeta (9/2) \zeta (7/2)
\frac{g(T)}{g_*^2}
\end{equation}
The cosmological equations can now be solved in the high energy
regime leading to the Weyl parameter on the brane
\begin{equation}
C(t) \sim \alpha a^4
\end{equation}
and
\begin{equation}
a(t)\sim t^{1/(4+\alpha)}
\end{equation}
Numerically $\alpha \sim 0.02$ in the standard model of particle
physics. The correction to the $t^{1/4}$ behaviour in the high
energy phase is tiny. Nevertheless, notice that the ratio
\begin{equation}
\epsilon_W= \frac{\rho_W}{\rho_{rad}} \to \frac{\alpha}{4}
\end{equation}
at low energy. 
The presence of some extra radiation energy density during
nucleosynthesis is highly constrained leading to
\begin{equation}
\epsilon_W\le 8. 10^{-2}
\end{equation}
which is marginally higher than $\alpha /4$. Forthcoming
constraints from CMB anisotropies will lead to a tightening of
this bound. This might open up the possibility of confronting
the prediction of brane models with experiment.

In the following we will present the case when radiation is not
radial.

\subsubsection{Gravitons bouncing off the brane}

So far it has been assumed that radiation is radial. This is a drastic
assumption. In particular this implies that gravitons cannot be
emitted and then captured by the brane. An analysis which
includes the effect of gravitons bouncing off the brane has been performed
in \cite{graviton2}. The gravitons emitted after a
particle--antiparticle annihilation on the brane are not all
radial. In particular, the energy--momentum tensor is not well
approximated by the form given in (\ref{mom}) but also contains
non-radial terms proportional to $\rho^2$. Moreover, the
gravitons may follow geodesics in the bulk which could lead to
subsequent collisions with the brane. As a result,
the non--conservation equation for the dark component depends on
the amount of radiation being emitted and bouncing off the brane. This has been
studied numerically. In particular the ratio $\epsilon_W$ has been
evaluated and is enhanced in comparison with the value, $\alpha /4$,
given in the previous section. Consequently, this tightens the bounds
coming from nucleosynthesis and CMB. Finally, the dynamical consequences of 
a non--constant Weyl parameter in radiating branes have been investigated 
in \cite{leeper}.

\subsection{Non-$Z_2$ Symmetric Branes}
Most brane world scenarios assume a $Z_2$ symmetry about our
brane, motivated by M-theory. However, a lot of recent models are
not directly derived from M-theory and the motivation for
maintaining the $Z_2$ symmetry is less clear. For example, 
multi-brane scenarios have been suggested which involve branes
that, although lying between two bulk space times with the same
cosmological constant, {\it do not possess a $Z_2$ symmetry of the
metric itself}~\cite{Ross}. In this subsection we investigate the
cosmological implications of relaxing the $Z_2$ symmetry about the
brane world embedded in AdS--Schwarzschild \cite{tye},\cite{anne2}.

When one does not impose a $Z_2$ symmetry about our brane world the
Friedmann equation on the brane acquires an extra term for the lack
of $Z_2$ symmetry. The method of deriving the Friedmann equation is
as outlined previously. However, the junction conditions are modified. 
Let us write \cite{anne2}
\begin{equation} \label{defn_d}
a'(0^+) = -a'(0^-) + d(t).
\end{equation}
where $d(t)$ is some function of time only and has yet to be determined.
$d(t)$ represents the asymmetry of the metric across the brane and the
fact that it is not identically zero for all time is what is referred
to as the brane being `non-$Z_2$ symmetric' in the rest of this review.
Calculating the Israel junction conditions one finds that
\begin{equation}\label{defnd}
d(t) \; = \; \frac{2F}{\rho_b a^3},
\end{equation}
where $F$ is an integration constant which, when non-zero,  dictates to
what extent the $Z_2$ symmetry is broken (in fact, it is related to the 
difference of the black hole masses on each sides of the branes, see 
\cite{tye}). This results in a modified
Friedmann equation
\begin{equation} \label{Fried_eqtn}
\left(
      \frac{\dot{a}}{a}
\right)^2 = \frac{\kappa^4 \rho_{\lambda}}{18} \rho  \; + \;
\frac{\kappa^4}{36}\rho^2 \; - \; \frac{k}{a^2} \; + \;
\frac{\mathcal{C}}{a^4} \; + \; \frac{F^2}{(\rho +
\rho_{\lambda})^2 a^8}.
\end{equation}
where $\rho_\lambda$ is the energy density associated to the brane tension.
So the absence of the $Z_2$ symmetry gives rise to an extra term in
the Friedmann equation. For a radiation
dominated Universe where $\rho = \gamma/a^4$, the extra term behaves
as $F^2/\gamma^2$ as $\rho \rightarrow \infty$.
This results in the
Hubble constant being approximately constant at very early times, resulting
in an inflation-like period. At late times the extra term behaves as
$(F\rho/\gamma\rho_{\lambda})^2$ as $\rho \rightarrow 0$, so
the effects are no longer significant and our solution reverts to that
analysed previously.

The effect of this extra term in the Friedmann equation has been analysed
in \cite{anne2}, where it was shown that the universe undergoes a period
of exponential expansion during a radiation dominated epoch. 
However, given that the universe must be described
by the usual Friedmann equation by the time of nucleosynthesis the
maximum expansion due to the F-term was found to be  $10^4$, which is
less than that needed to solve the standard cosmological problems.

\subsection{Some final remarks}
The Randall--Sundrum brane world is the simplest brane world model, which leads 
to a consistent cosmological framework. However, the model is related to deeper 
issues, such as the AdS--CFT correspondence (see e.g. \cite{gubser}-\cite{kaloper2}) 
and therefore to the holographic principle \cite{thooft}. We cannot go into 
the details here. Similarly, we do not cover inflation driven by the trace 
anomaly of the conformal field theory living on the brane 
(see e.g. \cite{hawking1}-\cite{tracelast}). These are interesting and important 
developments which give insights into gravity and the consequences for early universe 
cosmology.

\section{Including a Bulk Scalar Field}

In this section we are going to generalise the results found in
the previous section where the bulk was empty to a more realistic
case where the effects of a bulk scalar field are included. This
is motivated by supersymmetric extensions to the Randall--Sundrum
model, and of course is a more realistic limit of string theory
which automatically has the string dilaton field. The projective
approach considered previously, where one focuses on the dynamics
of the brane, can be extended to this case. Here one
studies both the projected Einstein {\it and} the Klein-Gordon
equation \cite{maedawands}, \cite{mennim}. Again, bulk effects do
not decouple, so the dynamics are not closed. As we will see, there are
now two objects representing the bulk back-reaction: the projected
Weyl tensor $E_{\mu\nu}$ and the loss parameter $\Delta\Phi_2$. In
the case of homogeneous and isotropic cosmology on the brane, the
projected Weyl tensor is determined entirely up to a dark
radiation term. Unfortunately, no information on the loss
parameter is available. This prevents a rigorous treatment of
brane cosmology in the projective approach. However, the inclusion
of a bulk scalar field opens up other avenues. As we will see
later in this section, the bulk scalar field can give rise to 
late--time acceleration of the universe, i.e. it could act as a
quintessence field, and also leads to a variation in Newton's constant.

Another route to analysing brane models is to study the motion of a
brane in a bulk space--time. This approach is successful in the
Randall-Sundrum case thanks to Birkhoff's theorem which dictates a
unique form for the metric in the bulk \cite{charmousis}. In the
case of a bulk scalar field, no such theorem is available. One has
to resort to various ansatze for particular classes of bulk and
brane scalar potentials (see e.g. \cite{kanti1}--\cite{kanti2}).
We discuss this in more detail in section 4.

\subsection{BPS Backgrounds}
\subsubsection{Properties of BPS Backgrounds}
As the physics of branes with bulk scalar fields is pretty
complicated, we will start with a particular example where both
the bulk and the brane dynamics are related and under control
\cite{brax1} (see also \cite{groje} and \cite{youm}). This is the
situation in the BPS case and is automatic in the supersymmetric
case. We specify the bulk Lagrangian as

\begin{equation}
S=\frac{1}{2\kappa_5^2}\int d^5x
\sqrt{-g_5}\left(R-\frac{3}{4}\left(\left(\partial \phi\right)^2
+V(\phi)\right)\right)
\end{equation}
where $V(\phi)$ is the bulk potential. The boundary action depends
on a brane potential $U_B(\phi)$
\begin{equation}
S_B=-\frac{3}{2\kappa_5^2}\int d^4x \sqrt{-g_4}U_B(\phi_0)
\end{equation}
where $U_B(\phi_0)$ is evaluated on the brane. The BPS backgrounds
arise as a special case of this general setting with a particular
relationship between the bulk and brane potentials. This relation
appears in the study of $N=2$ supergravity  with vector multiplets
in the bulk. The bulk potential is given by
\begin{equation}
V=\left(\frac{\partial W}{\partial\phi}\right)^2-W^2
\end{equation}
where $W(\phi)$ is the superpotential. The brane potential is
simply given by the superpotential
\begin{equation}
U_B=W
\end{equation}
We would like to mention that the last two relations have also been
used to generate bulk solutions without necessarily
imposing supersymmetry \cite{flanagan2},\cite{karch}. Notice that
the Randall--Sundrum case can be retrieved by putting $W=cst$.
Supergravity puts further constraints on the superpotential which
turns out to be of the exponential type \cite{brax1}
\begin{equation}
W=4k e^{\alpha \phi}
\end{equation}
with $\alpha=-1/\sqrt{12},1/\sqrt 3$. In the following we will
often choose this exponential potential with an arbitrary $\alpha$
as an example. The actual value of $\alpha$ does not play any role
and will be considered generic.

The bulk equations of motion comprise the Einstein equations and
the Klein-Gordon equation. In the BPS case and using the following
ansatz for the metric
\begin{equation}\label{scalarfieldmetric}
ds^2 = a(y)^2 \eta_{\mu\nu}dx^\mu dx^\nu + dy^2,
\end{equation}
these second order differential equations reduce to a system of
two first--order differential equations
\begin{eqnarray}
\frac{a'}{a}&=& -\frac{W}{4},\\\nonumber \phi'&=&\frac{\partial
W}{\partial\phi}.\nonumber
\end{eqnarray}
Notice that when $W=cst$ one easily retrieves the exponential
profile of the Randall-Sundrum model.

An interesting property of BPS systems can be deduced from the
study of the boundary conditions. The Israel junction conditions
reduce to
\begin{equation}
\frac{a'}{a}\vert_B= -\frac{W}{4}\vert_B
\end{equation}
and for the scalar field
\begin{equation}
\phi'\vert_B=\frac{\partial W}{\partial\phi}\vert_B
\end{equation}
This is the main property of BPS systems: the boundary conditions
coincide with the bulk equations, i.e. as soon as the bulk
equations are solved one can locate the BPS branes anywhere in
this background, there is no obstruction due to the boundary
conditions. In particular two-brane systems with two boundary BPS
branes admit {\it moduli} corresponding to massless deformations
of the background. They are identified with the positions of the
branes in the BPS background. We will come back to this later in
section 5.

Let us treat the example of the exponential superpotential. The
solution for the scale factor reads
\begin{equation}\label{aBPS}
a=(1-4k\alpha^2y)^{1/4\alpha^2},
\end{equation}
and the scalar field is given by
\begin{equation}\label{phiBPS}
\phi=-\frac{1}{\alpha}\ln(1-4k\alpha^2y).
\end{equation}
For $\alpha\to 0$, the bulk scalar field decouples and these
expressions reduce to the Randall-Sundrum case. Notice a new
feature here, namely the existence of singularities in the bulk,
corresponding to
\begin{equation}
a(y)\vert_{y_*}=0
\end{equation}
In order to analyse singularities it is convenient to use
conformal coordinates
\begin{equation}
du=\frac{dy}{a(y)}.
\end{equation}
In these coordinates light follows straight lines $u=\pm t$. It is
easy to see that the singularities fall into two categories
depending on $\alpha$. For $\alpha^2< 1/4$ the singularity is at
infinity $u_*=\infty$. This singularity is null and absorbs
incoming gravitons. For $\alpha^2>1/4$ the singularity is at
finite distance. It is time-like and not wave-regular, i.e. the
propagation of wave packets is not uniquely defined in the
vicinity of the singularity. For all these reasons these naked
singularities in the bulk are a major drawback of brane models
with bulk scalar fields \cite{braxsingularity}. In the two-brane
case the second brane has to sit in front of the naked
singularity. This is automatic in the supersymmetric case, as
required by conservation of global charges.

\subsubsection{de Sitter and anti de Sitter Branes}

Let us modify slightly the BPS setting by detuning the tension of
the BPS brane. This corresponds to adding or subtracting some
tension compared to the BPS case
\begin{equation}
U_B=TW
\end{equation}
where $T$ is a 
real number. Notice that this modification only
affects the boundary conditions; the bulk geometry and scalar
field are still solutions of the BPS equations of motion. In this
sort of situation, one can show that the brane does not stay
static. For the detuned case, the result is either a boosted brane
or a rotated brane. Defining by $u(x^\mu)$ the position of the
brane in conformal coordinates, one obtains
\begin{equation}
(\partial u)^2= \frac{1-T^2}{T^2}.
\end{equation}
The brane velocity vector $\partial_\mu u$ is of constant norm.
For $T>1$, the brane velocity vector is time-like and the brane
moves at constant speed. For $T<1$ the brane velocity vector is
space-like and the brane is rotated\cite{br1,br2}. Going back to a static brane,
we see that the bulk geometry and scalar field become $x^\mu$
dependent. In the next section we will find many more cases where
branes move in a static bulk or equivalently, a static brane
borders a non-static bulk.

Let us now conclude this section by studying the brane geometry
when $T>1$. In particular one can study the Friedmann equation for
the induced scale  factor
\begin{equation}
H^2=\frac{T^2-1}{16}W^2,
\end{equation}
where $W$ is evaluated on the brane. Of course we find
that cosmological solutions are only valid for $T>1$. Now in the
Randall-Sundrum case $W=4k$ leading to
\begin{equation}
H^2=(T^2-1)k^2.
\end{equation}
In the case $T>1$ the brane geometry is driven by a positive
cosmological constant. This is a de Sitter brane. When $T<1$ the
cosmological constant is negative, corresponding to an AdS brane.
We are going to generalise these results by considering the
projective approach to the brane dynamics.

\subsection{Bulk Scalar Fields and the Projective Approach}
\subsubsection{The Friedmann Equation}

We will first follow the traditional coordinate--dependent path.
This will allow us to derive the matter conservation equation, the
Klein-Gordon and the Friedmann equations on the brane. Then we
will concentrate on the more geometric formulation where the role
of the projected Weyl tensor will become transparent
\cite{brax2},\cite{vandebruck2}. Again, in this subsection we will
put $\kappa_5\equiv1$.

We consider a static brane that we choose to put at the origin
$y=0$. and impose $ b(0,t)=1$. This guarantees that the brane and
bulk expansion rates
\begin{equation}
4H=\partial_\tau \sqrt{-g}\vert_0,\ 3H_B=\partial_\tau
\sqrt{-g_4}\vert_0
\end{equation}
coincide. We have identified the brane cosmic time $d\tau
=ab\vert_0 dt$. We will denote by prime the normal derivative $
\partial_n=\frac{1}{ab}\vert_0 \partial_{y}
$. Moreover we now allow for some matter to be present on the
brane
\begin{equation}
\tau ^\mu_\nu\  ^{matter}=(-\rho_m,p_m,p_m,p_m).
\end{equation}
The bulk energy-momentum tensor reads
\begin{equation}
T_{ab}= \frac{3}{4}\left(\partial_a\phi\partial_b\phi\right)
-\frac{3}{8}g_{ab}\left(\left(\partial\phi\right)^2+V\right).
\end{equation}
The total matter density and pressure on the brane are given by
\begin{equation}
\rho= \rho_m+\frac{3}{2}U_B,\ p=p_m -\frac{3}{2}U_B.
\end{equation}
The boundary condition for the scalar field is unchanged by the
presence of matter on the brane.

The $(05)$ Einstein equation leads to matter conservation
\begin{equation}
\dot \rho_m=-3H(\rho_m+p_m).
\end{equation}
By restricting the $(55)$ component of the Einstein equations we
obtain
\begin{equation}
H^2=\frac{\rho^2}{36}-\frac{2}{3}Q-\frac{1}{9}E +\frac{\cal
\mu}{a^4}
\end{equation}
in units of $\kappa_5^2$. The last term is the dark radiation,
whose origin is similar to the Randall-Sundrum case. The quantities
$Q$ and $E$ satisfy the differential equations \cite{vandebruck1}
\begin{eqnarray}
\dot Q+4HQ&=&HT^5_5,\nonumber \\
\dot E + 4HE &=& -\rho T^0_5. \nonumber
\end{eqnarray}
These equations can be easily integrated to give
\begin{equation}
H^2= \frac{\rho_m^2}{36}+\frac{U_B\rho_m}{12}-\frac{1}{16a^4}\int
d\tau \frac{da^4}{d\tau}(\dot\phi^2-2U)-\frac{1}{12a^4}\int d\tau
a^4 \rho_m \frac{dU_B}{d\tau},
\end{equation}
up to a dark radiation term and we have identified
\begin{equation}
U=\frac{1}{2}\left(U_B^2-\left(\frac{\partial U_B}{\partial
\phi}\right)^2+V\right).
\end{equation}
This is  the Friedmann equation for a brane coupled to a bulk
scalar field. Notice that retarded effects springing from the
whole history of the brane and scalar field dynamics are present.
In the following section we will see that these retarded effects
come from the projected Weyl tensor. They arise from the exchange
between the brane and the bulk. Notice that Newton's constant
depends on the value of the bulk scalar field evaluated on the
brane ($\phi_0=\phi(t,y=0)$):
\begin{equation}
\frac{8\pi G_N(\phi_0)}{3} =\frac{\kappa_5^2 U_B(\phi_0)}{12}.
\end{equation}
On cosmological scales, time variation of the scalar field induces
a time variation of Newton's constant. This is highly constrained
experimentally \cite{chiba},\cite{uzan}, leading to tight
restrictions on the time dependence of the scalar field.

To get a feeling of the physics involved in the Friedmann
equation, it is convenient to assume that the scalar field is
evolving slowly on the scale of the variation of the scale factor.
Neglecting the evolution of Newton's constant, the Friedmann
equation reduces to
\begin{equation}
H^2= \frac{8\pi G_N(\phi)}{3}\rho_m +\frac{U}{8}-\frac{\dot
\phi^2}{16}
\end{equation}
Several comments are in order. First of all we have neglected the
contribution due to the $\rho_m^2$ term as we are considering
energy scales below the brane tension. Then the main effect of the
scalar field dynamics is to involve the  potential energy $U$ and
the kinetic energy $\dot \phi^2$. Although the potential energy
appears with a positive sign we find that the kinetic energy has a
negative sign. For an observer on the brane this looks like a
violation of unitarity. We will reanalyse this issue in section 5,
when considering the low energy effective action in four
dimensions and we will see that there is no unitarity problem in
this theory. The minus sign for the kinetic energy is due to the
fact that one does not work in the Einstein frame where Newton's
constant does not vary, a similar minus sign appears also in the
effective four--dimensional theory when working in the brane
frame.

The time dependence of the scalar field is determined by the
Klein-Gordon equation. The dynamics is completely specified by
\begin{equation}
\ddot\phi
+4H\dot\phi+\frac{1}{2}(\frac{1}{3}-\omega_m)\rho_m\frac{\partial
U_B}{\partial\phi}= -\frac{\partial U}{\partial\phi}
+\Delta\Phi_2,
\end{equation}
where $p_m=\omega_m\rho_m$. We have identified
\begin{equation}
\Delta\Phi_2=\phi''\vert_0-\frac{\partial U_B}{\partial
\phi}\frac{\partial^2 U_B}{\partial \phi^2}\vert_0.
\end{equation}
This cannot be set to zero and requires the knowledge of the
scalar field in the vicinity of the brane. When we discuss
cosmological solutions below, we will assume that this term is
negligible.

The evolution of the scalar field is driven by two effects. First
of all, the scalar field couples to the trace of the energy--momentum 
tensor via the gradient of $U_B$. Secondly, the field is
driven by the gradient of the potential $U$, which might not
necessarily vanish.

\subsubsection{The Friedmann equation vs the projected Weyl tensor}

We are now coming back to the origin of the non-trivial Friedmann
equation. Using the Gauss-Codazzi equation one can obtain the
Einstein equation on the brane \cite{maedawands},\cite{mennim}
\begin{equation}
\bar G_{ab}=
-\frac{3}{8}Uh_{ab}+\frac{U_B}{4}\tau_{ab}+\pi_{ab}+\frac{1}{2}
\partial_a \phi \partial_b\phi -\frac{5}{16}(\partial\phi)^2 h_{ab}-E_{ab}.
\end{equation}
Now the projected Weyl tensor can be determined in the homogeneous
and isotropic cosmology case. Indeed  only the $E_{00}$ component
is independent. Using the Bianchi identity $\bar D^a \bar
G_{ab}=0$ where $\bar D_a$ is the brane covariant derivative, one
obtains that
\begin{equation}
\dot E_{00}+4HE_{00}= \partial_{\tau}\left(\frac{3}{16}\dot
\phi^2+\frac{3}{8}U\right)+ \frac{3}{2}H\dot\phi^2 +\frac{\dot
U_B}{4}\rho_m
\end{equation}
leading to
\begin{equation}
E_{00}=\frac{1}{a^4}\int d\tau
a^4\left(\partial_{\tau}\left(\frac{3}{16}\dot \phi^2
+\frac{3}{8}U\right) + \frac{3}{2}H\dot\phi^2 +\frac{\dot
U_B}{4}\rho_m\right)
\end{equation}
Upon using
\begin{equation}
\bar G_{00}=3H^2
\end{equation}
one obtains the Friedmann equation. It is remarkable that the
retarded effects in the Friedmann equation all spring from the
projected Weyl tensor. Hence the projected Weyl tensor proves to
be much richer in the case of a bulk scalar field than in the
empty bulk case.

\subsubsection{Self--Tuning and Accelerated Cosmology}
The dynamics of the brane is not closed, it is an open system
continuously exchanging energy with the bulk. This exchange is
characterised by the dark radiation term and the loss parameter.
Both require a detailed knowledge of the bulk dynamics. This is of
course beyond the projective approach where only quantities on the
brane are evaluated. In the following we will {\it assume} that
the dark radiation term is absent and that the loss parameter is
negligible. Furthermore, we will be interested in the effects of a
bulk scalar field for late--time cosmology (i.e. well after
nucleosynthesis) and not in the case for inflation driven by a
bulk scalar field (see e.g. \cite{bscosfirst}-\cite{bscoslast}).

Let us consider the self--tuned scenario as a solution to the
cosmological constant problem. It corresponds to the BPS
superpotential with $\alpha=1$. In that case the potential $U=0$
for any value of the brane tension. The potential $U=0$ can be
interpreted as a vanishing of the brane cosmological constant. The
physical interpretation of the vanishing of the cosmological
constant is that the brane tension curves the fifth dimensional
space-time leaving a flat brane intact. Unfortunately, the
description of the bulk geometry in that case has shown that there
was a bulk singularity which needs to be hidden by a second brane
whose tension is fine tuned with the first brane tension. This
reintroduces a fine tuning in the putative solution to the
cosmological constant problem \cite{nilles}.

Let us generalise the self--tuned case to $\alpha\ne 1$, i.e.
$U_B=TW,\  T>1$ and $W$ is the exponential superpotential. The
resulting induced metric on the brane is of the FRW type with a
scale factor
\begin{equation}
a(t)= a_0 \left(\frac{t}{t_0}\right)^{1/3+ 1/6\alpha^2}
\end{equation}
leading to an acceleration parameter
\begin{equation}
q_0= \frac{6\alpha^2}{1+2\alpha^2}-1
\end{equation}
For the supergravity value $\alpha=-\frac{1}{\sqrt 12}$ this leads
to $q_0=-4/7$ and equation of state $\omega=-5/7$. This is in
coincidental agreement with the supernovae results. This model can
serve as a {\it brane quintessence model}
\cite{brax1},\cite{brax2}. We will comment on the drawbacks of
this model later. See also \cite{kiritsis} and \cite{tetradis} for
similar ideas.

\subsubsection{The brane cosmological eras}
Let us now consider the possible cosmological scenarios with a
bulk scalar field \cite{brax2},\cite{vandebruck2}. We assume  that
the potential energy of the scalar field $U$ is negligible
throughout the radiation and matter eras before serving as
quintessence in the recent past.

At very high energy, i.e. energies above the tension of the brane,
the non-conventional cosmology driven by the $\rho_m^2$ term in
the Friedmann equation is obtained. Assuming radiation domination,
the scale factor behaves like
\begin{equation}
a=a_0\left(\frac{t}{t_0}\right)^{1/4}
\end{equation}
and the scalar field
\begin{equation}
\phi=\phi_i+ \beta \ln\left(\frac{t}{t_0}\right)
\end{equation}
In the radiation--dominated era no modification is present,
provided
\begin{equation}
\phi=\phi_i
\end{equation}
which is a solution of the Klein-Gordon equation as the trace of
the energy--momentum tensor for radiation vanishes (together with a
decaying solution, which we have neglected). In the matter--dominated 
era the scalar field evolves due to the coupling to the
trace of the energy--momentum tensor. This has two consequences.
Firstly, the kinetic energy of the scalar field starts
contributing in the Friedmann equation. Secondly, the effective
Newton constant does not remain constant. The cosmological
evolution of Newton's constant is severely constrained since
nucleosynthesis \cite{chiba},\cite{uzan}. This restricts the
possible time variation of $\phi$.

In order to be more quantitative let us come back to the
exponential superpotential case with a detuning parameter $T$. The
time dependence of the scalar field and scale factor become
\begin{eqnarray}
\phi&=&\phi_1-\frac{8}{15}\alpha \ln\left(\frac{t}{t_e}\right)\nonumber \\
a&=&
a_e\left(\frac{t}{t_e}\right)^{\frac{2}{3}-\frac{8}{45}\alpha^2}\nonumber
\end{eqnarray}
where $t_e$ and $a_e$ are the time and scale factor at
matter-radiation equality. Notice the slight discrepancy of the
scale factor exponent with the standard model value of $2/3$. The
redshift dependence of the Newton constant is
\begin{equation}
\frac{G_N(z)}{G_N(z_e)}=\left(\frac{z+1}{z_e+1}\right)^{4\alpha^2/5}
\end{equation}
For the supergravity model with $\alpha=-\frac{1}{\sqrt 12}$ and
$z_e\sim 10^3$ this leads to a decrease by (roughly) 37\% since
nucleosynthesis. This is marginally compatible with experiments
\cite{chiba},\cite{uzan}.

Finally let us analyse the possibility of using the brane
potential energy of the scalar field $U$ as the source of
acceleration now. We have seen that when matter is negligible on
the brane, one can build brane quintessence models. We now require
that this occurs only in the recent past. As can be expected, this
leads to a fine--tuning problem as
\begin{equation}
M^4\sim \rho_c
\end{equation}
where $M^4=(T-1)\frac{3W}{2\kappa_5^2}$ is the amount of detuned
tension on the brane. Of course this is nothing but a
reformulation of the usual cosmological constant problem. Provided
one accepts this fine tuning, as in most quintessence models, the
exponential  model with $\alpha=-\frac{1}{\sqrt 12}$ is a
cosmologically consistent quintessence model with a five--dimensional
origin.

\subsection{Brief summary}
In this section we have explored the cosmology of a brane world model
with a bulk scalar field, as motivated by supersymmetry. This is a
complicated model, with more consequences than the 
Randall--Sundrum model. The main difference is that the gravitational
constant becomes time dependent. As such it has much in common
with scalar--tensor theories \cite{fujiibook}, but there are
important differences due to the projected Weyl tensor
$E_{\mu\nu}$ and its time evolution. The bulk scalar field can
play the role of the quintessence field, as discussed above, but
it could also play a role in an inflationary era in the very early
universe (see e.g. \cite{bscosfirst}-\cite{bscoslast}). In any
case, the cosmology of such a system is much richer and, because
of the variation of the gravitational constant, more constrained.
Similarly, it can leave a trace in the CMB anisotropies, which
will also constrain the parameters of the theory, and result in
the variation of fundamental constants. These results are
discussed in sections 5 and 6 of this review.

\section{Birkhoff Theorem and its Violations}

We have already mentioned the extended Birkhoff theorem in section $2$. It
states that for the case of a vacuum bulk space--time, the bulk is necessarily
static, in certain coordinates. A cosmologically evolving brane
is then moving in that space--time, whereas for an
observer confined to the brane the motion of the brane will be seen as an
expanding (or contracting) universe. In the case of a scalar field in the bulk,
a similar theorem is unfortunately not available, which makes the study of such
systems much more complicated. We will now discuss these issues in some detail,
following in particular \cite{stephen1} and \cite{stephen2}. As we will see 
later in this section, the violation of Birkhoff theorem in the presence of a 
bulk scalar field is due to the projected Weyl tensor.

\subsection{Motion in AdS-Schwarzschild Bulk}

We have already discussed the static background associated with
BPS configurations (including the Randall--Sundrum case) in the last
section. Here we will focus  on other backgrounds for
which one can integrate the bulk equations of motion.
Let us write the following ansatz for the metric
\begin{equation}\label{staticcoordinates}
ds^2 = -A^2(r)dt^2 + B^2(r) dr^2 + R^2(r)d\Sigma^2
\end{equation}
where $d\Sigma^2$ is the metric on the three--dimensional symmetric space of
curvature $q=0,\pm 1$. In general, the function $A,\ B$ and $R$
depend on the type of scalar field potential.
This is to be contrasted with the case of a negative bulk
cosmological constant where Birkhoff's theorem states that the
most general solution of the (bulk) Einstein equations is given
by $A^2=f$, $B^2=1/f$ and $R=r$ where
\begin{equation}
f(r)=q+\frac{r^2}{l^2}-\frac{\mu}{r^2}.
\end{equation}
We have denoted by $l=1/k=\sqrt{-6/(\Lambda_5\kappa_5^2)}$
the AdS scale and $\mu$ the black hole mass (see section 2).
This solution is the so--called AdS-Schwarzschild solution.

Let us now study the motion of a brane of tension $T/l$ in such a background.
The equation of motion is determined by the junction conditions.
 The resulting equation of motion for a boundary
brane with a $Z_2$ symmetry is
\begin{equation}
\left(\dot r^2+f(r)\right)^{1/2}=\frac{T}{l}r
\end{equation}
for a brane located at $r$ \cite{kraus}. Here $\dot r$ is the velocity of the
brane measured with respect to the proper time on the brane.
This leads to the following Friedmann equation
\begin{equation}
H^2 \equiv \left(\frac{\dot r}{r}\right)^2
= \frac{T^2-1}{l^2}-\frac{q}{r^2}+\frac{\mu}{r^4}.
\end{equation}
So the brane tension leads to an effective cosmological constant
$(T^2-1)/l^2$. The curvature gives the usual term familiar from
standard cosmology while the last term is the dark radiation term whose
origin springs from the presence of a black-hole in the bulk.
At late time the dark radiation term is negligible for an
expanding universe; we retrieve the cosmology of a FRW universe
with a non-vanishing cosmological constant. The case $T=1$
corresponds of course to the Randall--Sundrum case.

Notice that all the bulk physics in the absence of a bulk scalar field is 
captured by the unique choice of the static background solution. This solution 
is only parametrised by the $AdS$ curvature and the black hole mass. These two 
numbers are all that is needed to describe the motion of any brane in a static 
bulk. We will now contrast this situation with the case of a bulk scalar field.

\subsection{Violating Birkhoff Theorem}

We now turn to a general analysis of the brane motion in a static bulk with a 
bulk scalar field. To do that it is convenient to parametrise the bulk
metric slightly differently
\begin{equation}
ds^2= -f^2(r)h(r) dt^2 +\frac{dr^2}{h(r)}+ r^2d\Sigma^2_q.
\end{equation}
where $q=0,\ \pm 1 $ is the spatial curvature. 
Now, the Einstein equations lead to (redefining $\phi \to
\frac{\sqrt 3}{2\kappa_5}\phi$ and $V\to \frac{3}{8\kappa_5^2} V$)
\begin{eqnarray}
& &\frac{3}{r^2}\left(h+\frac{rh'}{2}-q\right)
= -{\kappa_5^2}\left( \frac{h\phi'^2}{2}+V\right) \label{eins}\\
& &\frac{3}{r^2}\left(h+\frac{rh'}{2}-q+\frac{hrf'}{f}\right)
= {\kappa_5^2}\left(\frac{h\phi'^2}{2}-V\right) \label{zwei}
\end{eqnarray}
and the Klein Gordon equation
\begin{equation}
h\phi''+\left(\frac{3h}{r}+\frac{h f'}{f}+h'\right)\phi'=\frac{dV}{d\phi}.
\end{equation}
Subtracting eq. (\ref{eins}) from (\ref{zwei}) and solving the resulting
differential equation, we obtain
\begin{equation}
f=\exp\left(\frac{\kappa_5^2}{3}\int dr r\phi'^2\right).
\end{equation}
This gives the link between the bulk scalar field and the bulk metric.

Another piece of information is obtained by
 evaluating the spatial trace of the projected
Weyl tensor. This is obtained by computing both the bulk Weyl
tensor and the vector normal to the moving brane
\begin{equation}
\frac{\mu}{r^4}\equiv -\frac{E^i_i}{3}=\frac{r}{4f^2}\left(\frac{h f^2}{r^2}\right)'+\frac{q}{2r^2}.
\end{equation}
This is the analogue of the dark radiation term for a general background.
The equations of motion can be cast in the form
\begin{eqnarray}
\mu'&=&-\frac{\kappa_5^2}{3}(\mu-\frac{kr^2}{2})r\phi'^2, \\
{\cal H}'+4\frac{\mu}{r^5}&=&-\frac{2\kappa_5^2}{3}({\cal H}-\frac{q}{r^2})r\phi'^2, \\
\kappa_5^2 V&=&6{\cal H}+\frac{3}{4}r{\cal H}'-3\frac{\mu}{r^4},
\end{eqnarray}
where we have defined
\begin{equation}
{\cal H}=\frac{q-h}{r^2}.
\end{equation}
This allows us to retrieve easily some of the previous solutions.

Choosing $\phi$ to be constant leads to $f=1$, $\mu$ is constant and
\begin{equation}
{\cal H}=-\frac{1}{l^2}+\frac{\mu}{r^4}
\end{equation}
This is the AdS-Schwarzschild solution as specified by Birkhoff theorem.

Another interesting case is obtained for spatially flat sections
 $q=0$
\begin{eqnarray}
\frac{\kappa_5^2}{3}r\phi'&=&-\frac{d\ln\mu}{d\phi}, \label{togetr} \\
d\left(\frac{\cal H}{\mu^2}\right)&=&\frac{1}{\mu}d\left(r^{-4}\right), \\
\frac{\kappa_5^2}{6}V&=&-\frac{3}{4\kappa_5^2}\frac{d\mu}{d\phi}\frac{d}{d\phi}
\left(\frac{\cal H}{\mu}\right) + {\cal H}\label{V}
\end{eqnarray}
In this form it is easy to see that the dynamics of the bulk are
completely integrable.
First of all the solutions depend on an arbitrary function $\mu (\phi)$
which determines the dynamics.
Notice that
\begin{equation}
f=\frac{\mu_0}{\mu}
\end{equation}
where $\mu_0$ is an arbitrary constant.
The radial coordinate $r$ is obtained by simple integration of eq. (\ref{togetr})
\begin{equation}
r=r_0e^{-\frac{\kappa_5^2}{3}\int \frac{d\phi}{d\ln\mu}d\phi}.
\end{equation}
Finally the rest of the metric follows from
\begin{equation}
h=-\frac{4\kappa_5^2}{3}r^2 \mu^2 \int d\phi \frac{d\phi}{d\mu}
e^{4\frac{\kappa_5^2}{3}\int \frac{d\phi}{d\ln\mu}d\phi}
\end{equation}
The potential $V$ then follows from (\ref{V}).
This is remarkable and  shows why Birkhoff's theorem
is not valid in the presence of a bulk scalar field.
It is due to the infinite number of choices for the dark radiation function
$\mu (\phi)$.
Any choice of this function leads to a new static background in the bulk.

Let us now discuss the cosmological consequences of this theorem.
The Friedmann equation for a cosmological brane in the static background 
specified by $\mu (\phi)$ is
\begin{equation}
H^2= {\cal H}+ \frac{\kappa_5^4}{36}\mu^2\rho^2
\end{equation}
where $H$ is the Hubble parameter on the brane in cosmic time.
In general the dynamics are extremely intricate. One can simplify and
retrieve standard cosmology by studying
the vicinity of a critical point $\frac{d\mu}{d\phi}=0$.
Parameterising
\begin{equation}
\mu= \frac{6A}{\kappa_5^2}+B\phi^2
\end{equation}
leads to the Friedmann equation
\begin{equation}
H^2=\frac{\kappa_5^4}{36}(\rho^2-\theta)\mu^2 +\frac{\mu}{a^4}+ o(a^{-4})
\end{equation}
Here $\theta$ is an arbitrary integration constant.
Notice that this is a small deviation from the Randall-Sundrum
case as the scalar field behaves as
\begin{equation}
\phi= r^{-B/A}
\end{equation}
and goes to zero at large distances. Hence, standard cosmology is
retrieved at low energy and long distance. Indeed, $\mu (\phi)$ becomes constant and putting $\theta=0$
one obtains the brane Friedmann equation with its characteristic $\rho^2$ term.

\section{Cosmology of a Two--Brane System}
In this section we will once more include an ingredient suggested
by particle physics theories, in particular M--theory. So far we
have assumed that there is only one brane in the whole
space--time. According to string theory, there should be at least
another brane in the bulk. Indeed, in heterotic M--theory these
branes are the boundaries of the bulk space--time
\cite{horavawitten}.  Another motivation is the hierarchy problem.
Randall and Sundrum proposed a two--brane model (one with positive
and one with negative tension), embedded in a five--dimensional
AdS space--time. In their scenario the standard model particles
would be confined on the {\it negative} tension brane. As they
have shown, in this case gravity is weak due to the warping of the
bulk space--time. However, as will become clear from the results in
this section, in order for this model to be consistent with
gravitational experiments, the interbrane distance has to be {\it
fixed} \cite{garriga}. This can be achieved, for example, with  a
bulk scalar field. As  shown in \cite{garriga} and \cite{chiba2},
gravity in the two--brane model of Randall-Sundrum is described by
a scalar--tensor theory, in which the interbrane distance, the
radion, plays the role of a  scalar field. Introduction of a bulk scalar 
field modifies the Brans--Dicke parameter (see \cite{brax3} and
\cite{brax4}) and introduces a second scalar field in the low--energy 
effective theory. Hence, in the case of two branes and a bulk scalar field,
the resulting theory at low energy is a bi--scalar--tensor theory
\cite{vandebruck3},\cite{cynolter}.

In the following we will investigate the cosmological consequences when
the distance between the branes is {\it not} fixed (for some aspects
not covered here see e.g. \cite{radcosfirst}-\cite{radcoslast}). Motivation for this
comes, for example, from a recent claim that the fine--structure constant
might slowly evolve with time \cite{barrow}.

\subsection{The low--energy effective action}
In order to understand the cosmology of the two--brane system, we
derive the low-energy effective action using the moduli space
approximation. The moduli space approximation gives the {\it
low--energy limit} effective action for the two--brane system,
i.e. for energies much smaller than the brane tensions. Later in this 
section we will compare the moduli space approximation to another method. 

In the static BPS solutions described in the section 3,
the brane positions can be chosen arbitrarily. In other words, they
are {\it moduli fields}. It is expected that by putting some
matter on the branes, these moduli fields become time-dependent, or,
if the matter is inhomogeneously distributed, space--time dependent.
Thus, the first approximation is to replace the brane positions
with space--time dependent functions. Furthermore, in order to allow
for the gravitational zero mode, we will replace the flat space--time
metric $\eta_{\mu\nu}$ with $g_{\mu\nu}(x^\alpha)$. We assume that
the evolution of these fields is slow, which means that we neglect
terms like $(\partial \phi)^3$ when constructing the low--energy
effective action.

As already mentioned, the moduli space approximation is only a good 
approximation at energies much less than the brane tension. {\it Thus, 
in the moduli space approximation we do
not recover the quadratic term in the Friedmann equation.}
(See \cite{vinet} for these issues.)
We are interested in the {\it late time} effects, where the corrections 
have to be small.

Replacing $\eta_{\mu\nu}$
with $g_{\mu\nu}(x^{\alpha})$ in (\ref{scalarfieldmetric}) and
collecting all the terms one finds from the five--dimensional action after 
an integration over $y$:
\begin{eqnarray}
S_{\rm MSA} &=& \int d^4 x \sqrt{-g_4}\left[ f(\phi,\sigma) {\cal R}^{(4)}
+ \frac{3}{4}a^2(\phi)\frac{U_B(\phi)}{\kappa_5^2}(\partial \phi)^2 \right.
\nonumber \\
&-& \left. \frac{3}{4} a^2(\sigma)\frac{U_B}{\kappa_5^2}(\sigma)(\partial \sigma)^2 \right].
\end{eqnarray}
with
\begin{equation}
f(\phi,\sigma) = \frac{1}{\kappa_5^2} \int^{\sigma}_{\phi} dy a^2 (y),
\end{equation}
with $a(y)$  given by (\ref{aBPS}).
The moduli $\phi$ and $\sigma$ represent the location
of the two branes.
Note that the kinetic term of the field $\phi$ has the wrong
sign. This is an artifact of the frame we use here. As we will see
below, it is possible to go to the Einstein frame with a simple
conformal transformation, in which the sign in front of the kinetic
term is correct for both fields.

In the following we will concentrate on the BPS system with exponential
superpotential from section 3. Let us redefine the fields according to
\begin{equation}
\tilde \phi^2 = \left(1 - 4k\alpha^2 \phi\right)^{2\beta}, \label{posia1}
\tilde \sigma^2 = \left(1-4k\alpha^2 \sigma\right)^{2\beta} \label{posia2},
\end{equation}
with $ \beta = \frac{2\alpha^2 + 1}{4\alpha^2}$;
and then
\begin{equation}
\tilde \phi = Q \cosh R, \label{posib1}  \
\tilde \sigma = Q \sinh R \label{posib2}.
\end{equation}
A conformal transformation $\tilde g_{\mu\nu} = Q^2 g_{\mu\nu}$
leads to the Einstein frame action:
\begin{eqnarray}
S_{\rm EF} &=& \frac{1}{2k\kappa^2_5(2\alpha^2 + 1)}
\int d^4x \sqrt{-g}\left[ {\cal R} -  \frac{12\alpha^2}{1+2\alpha^2}
\frac{(\partial Q)^2}{Q^2} \right. \nonumber \\
&-& \left. \frac{6}{2\alpha^2 + 1}(\partial R)^2\right].
\end{eqnarray}
Note that in this frame both fields have the correct sign in front of
the kinetic terms. For $\alpha \rightarrow 0$
(i.e.\ the Randall--Sundrum case) the $Q$--field decouples. This
reflects the fact that the bulk scalar field decouples, and the only
scalar degree of freedom is the distance between the branes.
One can read off the gravitational constant to be
\begin{equation}
16\pi G = 2k\kappa_5^2 (1+2\alpha^2).
\end{equation}

The matter sector of the action can be found easily: if matter lives
on the branes, it ``feels'' the induced metric. That is, the action
has the form
\begin{equation}\label{matter1}
S_m^{(1)} = S_m^{(1)}(\Psi_1,g^{B(1)}_{\mu\nu}) \hspace{0.5cm} {\rm and}
\hspace{0.5cm} S_m^{(2)} = S_m^{(2)}(\Psi_2,g^{B(2)}_{\mu\nu}),
\end{equation}
where $g^{B(i)}_{\mu\nu}$ denotes the induced metric on each branes.
In going to the Einstein frame one gets
\begin{equation}
S_m^{(1)} = S_m^{(1)}(\Psi_1,A^2(Q,R)g_{\mu\nu}) \hspace{0.5cm} {\rm and}
\hspace{0.5cm} S_m^{(2)} = S_m^{(2)}(\Psi_2,B^2(Q,R)g_{\mu\nu}),
\end{equation}
where matter now couples explicitly to the fields via the functions
$A$ and $B$, which we will give below (neglecting derivative interactions).
Note that we have assumed in (\ref{matter1}) that the matter fields on the 
branes do not {\it directly} couple to the bulk scalar field. 

The theory derived with the help of the moduli space approximation
has the form of a {\it multi--scalar}--tensor theory, in which
matter on both branes couples differently to the moduli fields.
We note that methods different from the moduli--space approximation
have been used in the literature in order to obtain the low--energy
effective action or the resulting field equations for a two--brane
system (see in particular
\cite{braneaction0}--\cite{braneaction4}). Qualitatively, the features
of the resulting theories agree with the moduli--space approximation
discussed above. We will come back to this point in section 5.4. 

In the following we will discuss observational constraints imposed
on the parameters of the theory.

\subsection{Observational constraints}
In order to constrain the theory, it is convenient to write the moduli
Lagrangian in the form
of a non-linear sigma model with kinetic terms
\begin{equation}
\gamma_{ij}\partial \phi^i\partial \phi^j,
\end{equation}
where $i=1,2$ labels the moduli $\phi^1=Q$ and $\phi^2=R$.
The sigma model couplings are here
\begin{equation}
\gamma_{QQ}= \frac{12\alpha^2}{1+2\alpha^2}\frac{1}{Q^2},\
\gamma_{RR}=\frac{6}{1+2\alpha^2}.
\end{equation}
Notice the potential danger of the $\alpha\to 0$ limit, the RS model,
where the coupling to $Q$ becomes very small. In an ordinary
Brans-Dicke theory with a single field, this would correspond to a
vanishing Brans-Dicke parameter which is ruled out
experimentally. Here we will see that the coupling to matter is such
that this is not the case. Indeed we can write the action expressing
the coupling to
ordinary matter on our brane
as
\begin{equation}
  A=a(\phi)f^{-1/2}(\phi,\sigma) , \  B=a(\sigma)f^{-1/2}(\phi,\sigma) ,
\end{equation}
where we have neglected the derivative interaction.

Let us introduce the parameters
\begin{equation}
\alpha_Q=\partial_Q \ln A,\ \alpha_R=\partial_R \ln A.
\end{equation}
We find that ($\lambda = 4/(1+2\alpha^2)$)
\begin{equation}
A=Q^{-\frac{\alpha^2\lambda}{2}}(\cosh R)^{\frac{\lambda}{4}},
\end{equation}
leading to
\begin{equation}
\alpha_Q= -\frac{\alpha^2\lambda}{2}\frac{1}{Q}, \alpha_R=\frac{\lambda
\tanh R}{4}.
\end{equation}
Observations constrain the parameter
\begin{equation}\label{bdparameter}
\theta=\gamma^{ij}\alpha_i\alpha_j
\end{equation}
to be less than $10^{-5}$ \cite{gilles} today. We obtain therefore a
bound on
\begin{equation}
\theta= \frac{1}{3}\frac{\alpha^2}{1+2\alpha^2}+ \frac{\tanh^2
R}{6(1+2\alpha^2)}.
\end{equation}
The bound implies that
\begin{equation}
\alpha \leq 10^{-2},\ R\leq 0.2
\end{equation}
today. The smallness of $\alpha$ indicates a strongly warped
bulk geometry such as an Anti--de Sitter space--time.  In the case
$\alpha=0$, we can easily interpret the bound on $R$.  Indeed in that
case
\begin{equation}
\tanh R = e^{-k(\sigma -\phi)},
\end{equation}
i.e. this is nothing but the exponential of the radion field measuring
the distance between the branes. Thus we find that
gravity experiments require the branes to be sufficiently far apart.
When $\alpha\ne 0$ but small, one way of obtaining a small value of $R$
is for the hidden brane to become close to the would-be singularity
where $a(\sigma)=0$.

Constraints on the parameters also arise from nucleosynthesis. Nucleosynthesis
constrains the effective number of relativistic degrees of freedom at this
epoch. In brane worlds the energy conservation equation implies
\begin{equation}
\rho a^3\ne const. ,
\end{equation}
resulting in a different expansion rate than that given by
general relativity, giving rise to constraints on the parameters.
One finds $\alpha \le 0.1$ and $R \le 0.4$ at the time of nucleosynthesis.

We would like to mention that the parameter $\theta$ can be calculated
also for matter on the negative tension brane. Then, following the
same calculations as above, it can be seen that the observational
constraint for $\theta$ {\it cannot} be satisfied. Thus, if the
standard model particles are confined on the negative tension brane,
{\it the moduli have necessarily to be stabilised.} In the following
we will assume that the standard model particles are confined on the
positive tension brane and study the cosmological evolution of
the moduli fields.

\subsection{Cosmological implications}
The discussion in the last subsection raises an important question:
the parameter $\alpha$ has to be chosen rather small for
the theory to be consistent with observations. Similarly the field $R$
has to be small too. The field $R$ is dynamical and one would
like to know if the cosmological evolution drives the field $R$ to
small values such that it is consistent with the observations today.
Otherwise  are there natural initial conditions for the field
$R$? In the following we study the cosmological evolution of the system in order to
answer these questions.

\subsubsection{Cosmological attractor solutions}

The field equations for a homogeneous and isotropic universe can be
obtained from the action. The Friedmann
equation reads
\begin{equation}\label{Friedmann}
H^2 = \frac{8 \pi G}{3} \left(\rho_1 + \rho_2 + V_{\rm eff}
+ W_{\rm eff} \right) + \frac{2\alpha^2}{1 + 2\alpha^2} \dot\phi^2
+ \frac{1}{1+2\alpha^2} \dot R^2.
\end{equation}
where we have defined $Q = \exp \phi$.
The field equations for $R$ and $\phi$ read
\begin{eqnarray}
\ddot R + 3 H \dot R &=& - 8 \pi G \frac{1+2\alpha^2}{6}\left[
\frac{\partial V_{\rm eff}}{\partial R} +
\frac{\partial W_{\rm eff}}{\partial R} \right. \nonumber \\
&+& \left. \alpha_R^{(1)} (\rho_1 - 3p_1) +
\alpha_R^{(2)} (\rho_2 - 3p_2) \right] \label{Rcos}
\end{eqnarray}
\begin{eqnarray}
\ddot \phi + 3 H \dot \phi &=&
-8 \pi G \frac{1+2\alpha^2}{12 \alpha^2} \left[
\frac{\partial V_{\rm eff}}{\partial \phi} +
\frac{\partial W_{\rm eff}}{\partial \phi} \right. \nonumber \\
&+& \left. \alpha_\phi^{(1)} (\rho_1 - 3p_1) +
\alpha_\phi^{(2)} (\rho_2 - 3p_2) \right].\label{Qcos}
\end{eqnarray}
The coupling parameters are given by
\begin{eqnarray}
\alpha_\phi^{(1)} &=& -\frac{2\alpha^2}{1+2\alpha^2}, \hspace{0.5cm}
\alpha_\phi^{(2)} = -\frac{2\alpha^2}{1+2\alpha^2}, \label{coupling1} \\
\alpha_R^{(1)} &=& \frac{\tanh R}{1+2\alpha^2}, \hspace{0.5cm}
\alpha_R^{(2)} = \frac{(\tanh R)^{-1}}{1+2\alpha^2}. \label{coupling2}
\end{eqnarray}
We have included matter on both branes as well as potentials
$V_{\rm eff}$ and $W_{\rm eff}$.
We now concentrate on the case where matter is only on our brane.

We will discuss the inflationary epoch in more detail below.
In the radiation--dominated epoch the trace of the energy--momentum tensor 
vanishes, so that
$R$ and $\phi$ quickly become constant. The scale factor scales like
$a(t) \propto t^{1/2}$.

In the matter--dominated era the solution to these equations is given by
\begin{equation}
\rho_1 = \rho_e\left(\frac{a}{a_e}\right)^{-3-2\alpha^2/3}
,a=a_e\left(\frac{t}{t_e}\right)^{2/3-4\alpha^2/27}
\end{equation}
together with
\begin{eqnarray}
\phi &=& \phi_e+\frac{1}{3}\ln\frac{a}{a_e}, {\mbox{\vspace{0.5cm}}}
R = R_0\left(\frac{t}{t_e}\right)^{-1/3}
+ R_1\left(\frac{t}{t_e}\right)^{-2/3},
\end{eqnarray}
as soon as $t\gg t_e$. Note that $R$ indeed decays. This implies
that small values of $R$ compatible with gravitational
experiments are favoured by the cosmological evolution. Note, however,
that the size of $R$ in the early universe is constrained by
nucleosynthesis as well as by the CMB anisotropies.
A large discrepancy between the values of $R$ during
nucleosynthesis and now induces a variation of the particle
masses, or equivalently Newton's constant, which is excluded experimentally.
One can show that
by putting matter on the negative tension brane as well, the
field $R$ evolves even faster to zero \cite{vandebruck3}. This behaviour is reminiscent
of the attractor solution in scalar--tensor theories \cite{damour}.

In the five--dimensional picture the fact that $R$ is driven to small
values means that the negative tension brane is driven towards the
bulk singularity. In fact, solving the equations numerically for more general
cases suggest that $R$ can even be negative, which is, in the
five--dimensional description, meaningless as the negative tension
brane would move through the bulk singularity.
Thus, in order to make any further
progress, one has to understand the bulk singularity
better\footnote{For $\alpha=0$ the theory is equivalent to
the Randall--Sundrum model. In this case the bulk singularity is
shifted towards the Anti--de Sitter boundary.}. Of
course, one could simply assume that the negative tension brane
is destroyed when it hits the singularity. A more interesting
alternative would be if the brane is repelled instead. It was
speculated that this could be described by some effective potential
in the low-energy effective action \cite{vandebruck3}.

\subsubsection{Boundary inflation driven by an inflaton field on the brane}
Let us consider now an inflationary epoch, driven by an inflaton field
embedded on the positive tension brane \cite{boundaryinflation},\cite{ashcroft2}. 
The action for the inflaton field $\chi$ is,
in the Einstein frame,
\begin{equation}
S_{inflaton} = \int \sqrt{-g} d^4 x \left[ -\frac{1}{2} A^2(\phi,R)
\left(\partial \chi \right)^2 - A^4(\phi,R) V_0 \chi^n \right],
\end{equation}
where we have assumed a typical chaotic inflationary potential for the inflaton
field. Explicitly, the potential reads
\begin{equation}\label{potential}
V(\chi,\phi,R) = V_0 \exp\left(-\frac{8\alpha^2}{1+2\alpha^2}\phi\right)
\left( \cosh R \right)^{\frac{4}{1+2\alpha^2}} \chi^n.
\end{equation}
It can be shown that the field $R$ is driven towards zero during the inflationary epoch. The
reason is the same as in the matter era: the trace of the energy--momentum tensor of the
inflaton field is not zero. In fact, it can be shown that $R$ decays quickly to zero according
to
\begin{equation}
R(t) \propto \ln \left[ \frac{1+\exp(-4ct)}{1-\exp(-4ct)} \right],
\end{equation}
where $c$ is a constant depending on the initial values of $\chi$ and $\phi$ and the
energy scale $V_0$. In any case, $R$ decays quickly to zero and its
energy density becomes negligible. We assume in the following that
$R$ does not play a role in the last 60 e--folds of the inflationary era. We
are then left with an inflationary scenario with two scalar fields.
It is clear from eq.
(\ref{potential}) that the effective energy scale
\begin{equation}
V_{\rm eff} = V_0 \exp\left(-\frac{8\alpha^2}{1+2\alpha^2}\phi\right)
\end{equation}
depends on the evolution of $\phi$. For example, for $n=2$, the mass of the 
inflaton is not constant, but rather depends on the evolution of $\phi$, 
since we consider the problem in the Einstein frame in which the 
Planck mass is constant, but the masses of particles vary. This has the effect 
that the background solutions are slightly modified relative to
the case in General Relativity.

The perturbations in this scenario were discussed in \cite{ashcroft2} for 
the cases of $n=2$ and $n=4$. In the case of $n=2$ and $\alpha = 0.01$, 
entropy perturbations do not play a role at all. The amplitude of the
spectrum of entropy perturbations is around one percent that of the 
curvature perturbation. Similar conclusions hold for the case of $n=4$. In 
figure 1 we plot the spectral index of the curvature perturbation $n_R$ 
versus $r_T = P_T/16 P_R$ ($P_T$ is the tensor power spectrum and
$P_R$ is the curvature perturbation power spectrum) for different cases of 
$\alpha$. As can be seen, increasing $\alpha$ results in an increase of the 
tensor perturbation relative to the curvature perturbation and in a smaller 
value of $n_R$. The potential with $n=4$ is already under pressure 
observationally, and increasing $\alpha$ makes this worse. At 
least for the potentials considered here, a large coupling parameter 
$\alpha$ is not desirable.

\begin{figure}[!ht]
\centering
\includegraphics[width=14cm,height=11cm]{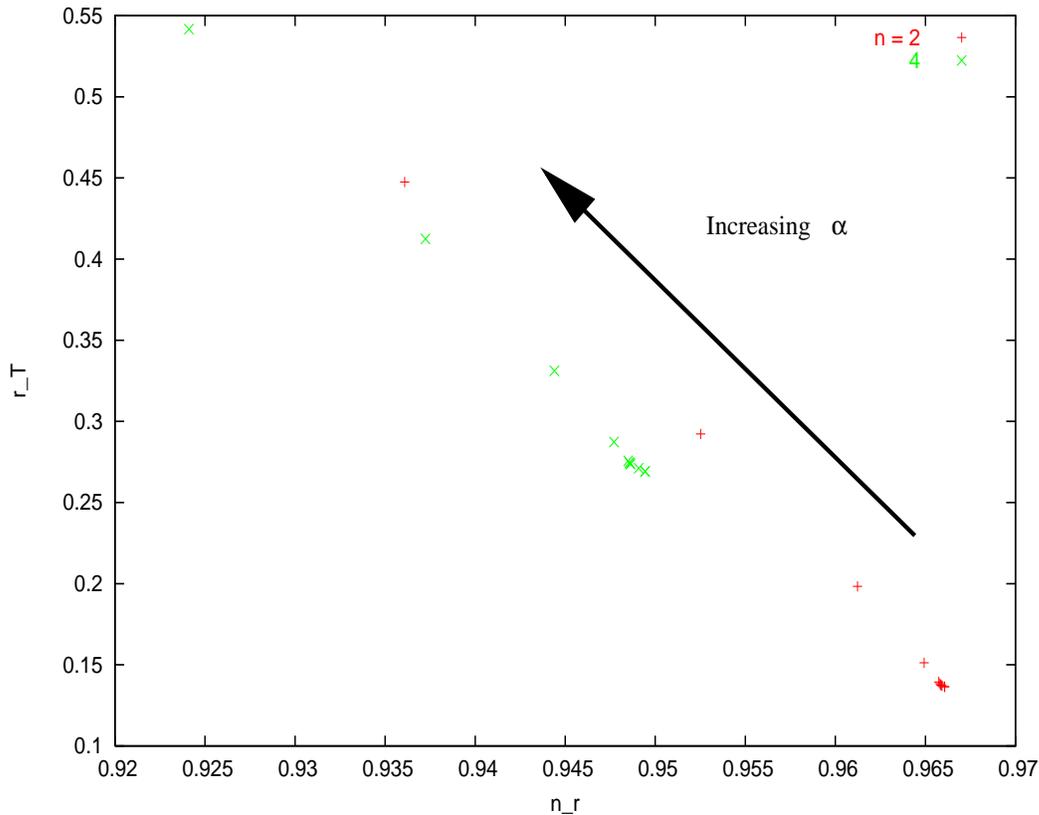}
\caption{As  can be seen, increasing $\alpha$ results in a
decrease of $n_R$, the power law index of the curvature
perturbation power spectrum and in an increase of $r_T= P_T/16
P_R$. Taken from \cite{ashcroft2}.}
\end{figure}

\subsubsection{Variation of constants}
In a realistic brane world model, i.e. in a model which is based
on a fundamental theory such as superstring theory, it is expected
that the moduli fields couple to all forms of matter embedded on the
brane \cite{varyour}. At the classical level, this automatically leads to the
conclusion that gauge coupling parameters, such as the fine
structure constant, are not fundamental parameters but depend on
the vacuum expectation values of the moduli fields themselves.
Quantum mechanically, an additional dependence of the couplings to
the moduli fields is generated by a conformal anomaly when
switching between the Jordan and the Einstein frame \cite{fujiibook}. It can,
however, be shown that one--loop corrections cancel and the
variation of constants is generated only by directly coupling the
gauge fields to the bulk scalar and having moduli--dependent Yukawa
couplings. In the following we assume that the Yukawa couplings do
not depend on the moduli fields. The expression for $\delta
\alpha_{EM} / \alpha_{EM}$ is, in general, a complicated function
of both $R$ and $\phi$. However, since $R$ rapidly decays to zero, we
can neglect the variation of $\alpha_{EM}$ due to that of
$R$. Then, using only the variation of $\phi$, one can find that
the variation of $\alpha_{EM}$ as a function of $z$ during the
matter dominated epoch reads
\begin{equation}
\frac{\delta \alpha_{EM}}{\alpha_{EM}} =
- \alpha_{EM}\frac{16\alpha\beta}{1+2\alpha^2}\ln \left( 1+z \right).
\end{equation}
In this equation, $\beta$ is a (constant) parameter, describing the coupling
of photons to the bulk scalar field (this parameter is similar to $\alpha$).

\begin{figure}[!ht]\label{varying}
\centering
\includegraphics[width=13cm,height=11cm]{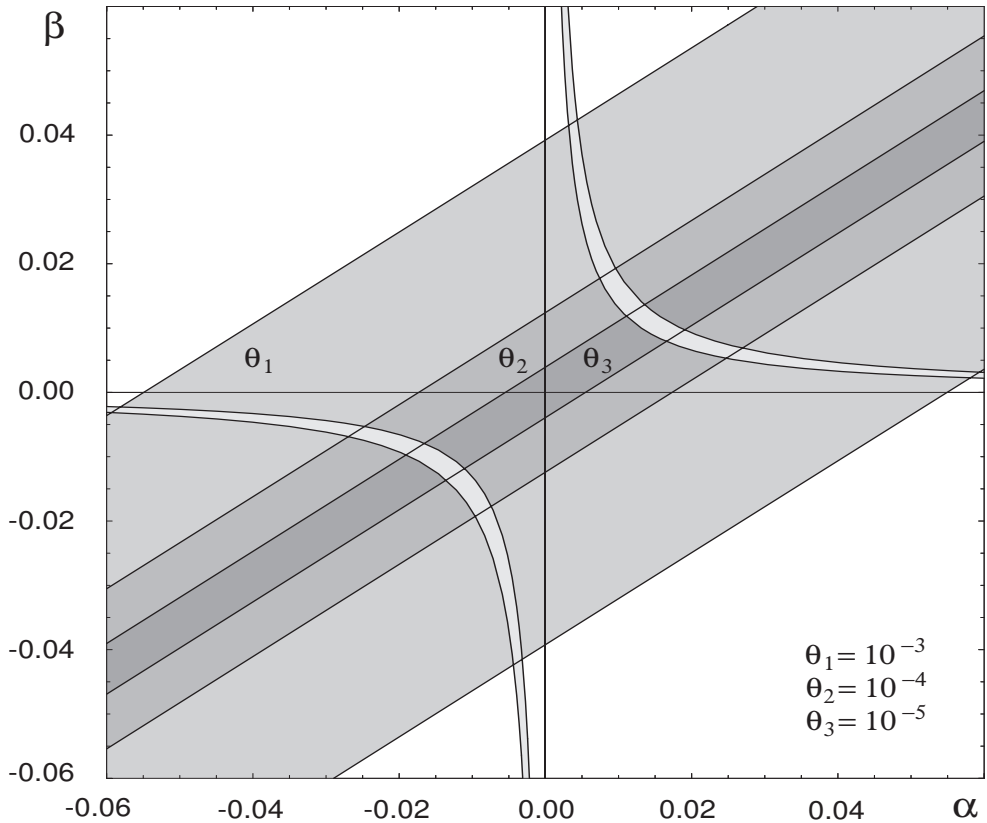}
\caption{Constraints on the parameter $\alpha$ and $\beta$, using the
results from quasar spectra (hyperbolic curves) and different values
for $\theta$. It is interesting to note that the observations, taken
at face value, imply that $\alpha$ and $\beta$ are of the same order. 
Taken from \cite{palma}.}
\end{figure}

In any case, it is a combination of $\alpha$ and $\beta$ which determines the
amplitude of the variation of $\alpha_{EM}$. We can now combine two 
observations which constrain both $\alpha$ and $\beta$: firstly, we can take 
the reported variation of $\alpha_{EM}$ for granted and fit the evolution 
of $\alpha_{EM}$
above to the data. Secondly, the parameter $\theta$ defined in 
(\ref{bdparameter}), also depends on $\beta$, since the masses of baryons are
directly proportional to the confinement
scale $\Lambda_{QCD}$, which depends on the moduli fields in the Einstein frame,
like all energy scales\footnote{We are considering the chiral limit here.}.
The analysis was carried out in \cite{palma} and the
result is shown in figure 2  under different assumptions for the value of
$\theta$.

It is interesting to note that these observations indicate  that
$\beta \approx \alpha$, which are {\it a priori} independent parameters.

\subsubsection{Open questions}
In this subsection, we have discussed some of the cosmological
consequences of brane world moduli. It is clearly a problem for
the theory that the negative tension brane collapses during the
cosmological evolution, although this attractor solution makes the
theory more compatible with observations. Nevertheless, theories
of this kind suffer from the existence of singularities and it is
conceivable that stringy effects will ``smooth out'' this
singularity. The cosmological implications of this might be
interesting in their own right.

Another potentially interesting question is how Kaluza--Klein
corrections affect the above results (see \cite{kaluza-klein} for 
an approach to include Kaluza--Klein corrections). Although it is unlikely that
these corrections prevent the collapse of the negative tension brane,
one can imagine that they influence the time taken 
for this to happen, influencing the dependence of $R(t)$ on $t$.

\subsection{Moduli versus Projective Approaches}

So far we have considered three different approaches to the brane
dynamics. The first one consists in solving the five dimensional
equations exactly (see Section 4). This is only possible in a few cases,
especially when no matter is present on the branes. The second one
is the projective approach (see Section 2 and Section 3). In that case
one writes down the four--dimensional equations on the branes.
In general one cannot find a closed system of equations.
Indeed the projected Weyl tensor is calculable only in an isotropic FRW
universe and reduces to the dark radiation term. In the case of a bulk
scalar field we have seen that the second derivative of the scalar field
with respect to the bulk coordinate is unconstrained from the point of
view of the brane. Thus, the full dynamics of the bulk scalar field is
not understood. Finally, we have presented the moduli space approximation
in the case of a two--brane system, whereby the whole dynamics have been
reduced to a four--dimensional description involving the low--energy degrees
of freedom.

In the following we will show how the projective approach and the moduli space
approximation are linked. This is only the case in the two--brane system.
In the one--brane system where the bulk is infinite, the calculation of the
projected Weyl tensor is an open  problem.

We will restrict the analysis to the Randall--Sundrum case, following \cite{braneaction1}. 
In that case the low energy degrees of freedom are the graviton and the radion. It is 
convenient to choose a gauge where the radion appears explicitly in the five dimensional metric
\begin{equation}
ds^2= e^{2\phi}dy^2 +g_{\mu\nu}(y,x^\mu) dx^\mu dx^\nu
\end{equation}
and the branes are placed at $y=0$ and $y=\rho$. In following we
will consider that $\phi(x^\mu)$ is a function in four dimensions whereas
$g_{\mu\nu}(y,x^\mu)$ depends on all five--dimensional coordinates.
Decomposing the extrinsic curvature
\begin{equation}
e^{-\phi}K_{\mu\nu}= \Sigma_{\mu\nu}+\frac{Q}{4}g_{\mu\nu}
\end{equation}
the junction conditions become
\begin{equation}
\Sigma_\nu^\mu -\frac{3}{4} \delta^\mu_\nu Q\vert_0=
\frac{\kappa_5^2}{2}\left(- \lambda_1 \delta^\mu_\nu+T^{(1)\mu}{\nu}\right)
\end{equation}
on the first brane of tension $\lambda_1$ and carrying the energy
momentum tensor $T^{(1),\mu}_{\nu}$. Together with the Einstein
equations, these junction conditions  specify the dynamics of the
system.

As we have already mentioned several times, we will be interested
in the low--energy limit of the brane system. This is the regime
where effects relevant for the CMB physics take place. High energy
effects would imply a modification of the early Universe features
which might leave an imprint on the primordial fluctuations. At
low energy on the branes, it is legitimate to use a derivative
expansion as long as typical length scales $L$ on the branes are
much larger than the curvature of the bulk $l=1/k$. Using
dimensional analysis the Hubble rate on the brane is $H=
O(L^{-1})$. The Friedmann equation when $\rho \ll \lambda$ leads to
$\rho =O(L^{-2}\kappa_4^{-2})$. Using $\lambda
=O(l^{-2}\kappa_4^{-2})$, this implies that
\begin{equation}
\frac{\rho}{\lambda}=  O\left(\left(\frac{l}{L}\right)^2\right)
\end{equation}
which is much smaller than one. Hence energy densities on the
brane are much lower than the brane tension. Within this
approximation, the bulk metric can be expanded
\begin{equation}
g_{\mu\nu}(y,x^\mu)= a^2(y)[ g_{\mu\nu}(x^\mu)+ g^{(1)}_{\mu\nu}+\dots]
\end{equation}
where at each order the tensor $g^{(n)}_{\mu\nu}$ involves only $(2n)$ derivatives on the brane, i.e. corresponds to terms of order $O((l/L)^{2n})$.
At each order in this approximation scheme, the function $g^{(n)}_{\mu\nu}(y,x^{\mu})$ of $y$ is valid throughout the bulk. Hence it encapsulates non-local effects due to the presence of both branes.

In the following we will analyse the zeroth and first order terms of the derivative expansion.
At zeroth order, we obtain the equations governing the Randall-Sundrum model
in the absence of matter on the branes.
At first order in the derivative expansion, the components 
of the extrinsic curvature tensor 
are given by
\begin{eqnarray}
\Sigma^{(1)\mu}_\nu&=&\frac{l}{a^2}[\frac{1}{2}(R^\mu_\nu
-\frac{1}{4} \delta^\mu_\nu R)+ \frac{ye^\phi}{l}(D^\mu D_\nu \phi
-\frac{1}{4}\delta^\mu_\nu D^2\phi) \nonumber \\ 
&+&(\frac{y^2 e^{2\phi}}{l^2} +\frac{ye^{\phi}}{l})(D^\mu\phi D_\nu \phi-
\frac{1}{4}\delta^\mu_\nu D^2\phi ) + \frac{l}{2}
\frac{E^{\mu}_{\nu}}{a^4}
\end{eqnarray}
where covariant  derivatives are taken with respect to the zeroth order metric
$g_{\mu\nu}(x^\mu)$.  Similarly the trace part reads
\begin{equation}
Q^{(1)}= \frac{l}{a^2}[\frac{R}{6} +\frac{ye^\phi}{l}(D^2\phi +
D^\alpha \phi D_\alpha \phi) -\frac{y^2e^{2\phi}}{l^2} D^\alpha \phi D_\alpha \phi]
\end{equation}
The Einstein equations lead to  the first order metric being a
function of the zeroth order terms
\begin{eqnarray}
g_{\mu\nu}^{(1)}&=& -\frac{l^2}{2}(\frac{1}{a^2}-1)(R_{\mu\nu}
-\frac{1}{6} g_{\mu\nu} R) \nonumber
\\ &+&\frac{l^2}{2}(\frac{1}{a^2}-1
-2\frac{2ye^\phi}{l}\frac{1}{a^2})(D_\mu D_\nu
\phi+\frac{1}{2}g_{\mu\nu}D^\alpha \phi D_{\alpha}\phi -\frac{y^2
e^{2\phi}}{a^2}(D_\mu\phi D_\nu\phi \nonumber \\ &-&\frac{1}{2}
g_{\mu\nu} D^\alpha \phi D_{\alpha}\phi)-(\frac{1}{a^4}-1)
E_{\mu\nu}
\end{eqnarray}
Notice that the metric $g_{\mu\nu}$ and $\phi$ are not fixed yet.
The equations of motion for gravity and $\phi$ follow from the
junction conditions. On the first brane, they can be cast in the
form of Einstein equations for the induced metric $g_{\mu\nu}$
\begin{equation}
G^{\mu}_{\nu}= \frac{\kappa_5^2}{l} T^{(1)\mu}_{\nu} +E^{\mu}_\nu
\end{equation}
where $G^\mu_\nu$ is the Einstein tensor on the first brane. Of
course this is nothing but the expected Einstein equation on the
first brane. On the second brane, the induced metric is
conformally related to the induced metric on the first one
$f_{\mu\nu}= \Omega^2 g_{\mu\nu}$ where $\Omega= \exp(-e^\phi)$
and
\begin{equation}
G^{\mu}_{\nu}= -\frac{\kappa_5^2}{l} T^{(2)\mu}_{\nu}
+\frac{E^{\mu}_\nu}{\Omega^4}
\end{equation}
Notice that the projected Weyl tensor appears in both Einstein
equations on the branes. Using the explicit relation between
$f_{\mu\nu}$ and $g_{\mu\nu}$ one can express the Einstein tensor
on the second brane in terms of the Einstein tensor on the first
brane and tensors depending on the conformal factor $\Omega$. This
allows one to obtain  $E^\mu_\nu$ from both equations
\begin{eqnarray}\label{weylproj}
E^\mu_\nu&=&
-\frac{\kappa_5^2}{l}\frac{1-\Psi}{\Psi}(T^{(1)\mu}_\nu + (1-\Psi)
T^{(2) \mu}_\nu) -\frac{1}{\psi}[(D^\mu D_\nu\Psi -\delta^\mu_\nu
D^2 \Psi) \nonumber \\ &+&\frac{w(\Psi)}{\Psi}(D^\mu \Psi D_\nu
\Psi -\frac{1}{2} \delta^\mu_\nu D^\alpha \Psi D_\alpha
\Psi)]
\end{eqnarray}
where $\Psi=1-\Omega^2$ and
\begin{equation}
w(\Psi)= \frac{3\Psi}{2(1-\Psi)}.
\end{equation}
Using the tracelessness of $E^\mu_\nu$, one obtains the field
equation governing the dynamics of $\Psi$. Moreover, now that we
have determined $E^\mu_\nu$, the Einstein equations on the first
brane lead to a closed set of equations. These equations of motion
are equivalent to the ones obtained with the moduli approximation
provided one identifies
\begin{equation}
\Psi=\frac{1}{\cosh^2 (R)}
\end{equation}
This shows the equivalence between the moduli space approach and
the projective approach. The solutions of the moduli space
equations of motion allow one to determine both $E^\mu_\nu$ and the
first order metric $g^{(1)}_{\mu\nu}$. Of course, this result
implies that the equations of motion of the moduli space approach
give an accurate description of the brane dynamics (at lowest order 
in the projective approach). For other work in this direction, we 
refer to \cite{braneaction0}-\cite{braneaction4} and 
\cite{proapproachini}-\cite{proapproachfini}.

Having shown the relationship between the projective approach and moduli
space approximation we now discuss cosmological perturbations and the 
effect the brane world moduli have on the CMB anisotropies.

\section{Cosmological Perturbations}
In this section we turn our attention to cosmological perturbations in
brane world scenarios. It might be that extra dimensions
have an important impact on the evolution (and maybe on the
generation) of cosmological perturbations. Therefore, cosmological
observations might be able to constrain extra--dimensional models, 
particularly with the advent of high precision data.

There has been a huge number of publications related to cosmological
perturbations in brane world scenarios, see 
\cite{perturbationsfirst}-\cite{perturbationslast}.
In this review we just touch the topic briefly and
describe some important results, since the details are very technical.
Instead we refer for more details to the reviews of brane world
perturbations in \cite{nathaliereview} and \cite{royreview}.

\subsection{Cosmological Perturbations: Methods}
Generally speaking, there are two points of view one can take.
The first is the point of view from a brane observer. In this
approach one starts with the Einstein equations on the brane
and considers perturbations of them. As we have seen, the projected
Weyl tensor $E_{\mu\nu}$ describes the influence of the bulk.
We will see below that the field equations on the brane do not close
and hence this approach is incomplete.

The second approach takes the point of view from the bulk. Here, one
can make two choices of coordinates: (a) the first choice of coordinates
assumes the brane to be at rest and straight, even in the presence of
perturbations (Gaussian normal coordinate system).
It turns out that in this coordinate system the junction conditions
are easy to implement. However, the bulk is non--static and, more
annoyingly, a coordinate singularity appears at finite distance
from the brane \cite{binetruy2}. One way to regulate this singularity is to 
introduce
a ``regulator brane'', on which boundary conditions have to be chosen. 
This method was used in \cite{koyamacmb} and we will come back to this later.
In addition to the singularity, apart from the case of a de--Sitter or 
Minkowski brane, the field equations are non--separable. 
(b) Because of the difficulties in the Gaussian normal coordinate system,
one may choose coordinates in which the bulk metric take a simple
form. Choosing a proper gauge in the bulk, one can then find the general
solution to the bulk equations \cite{nathalie}. The price to be paid when
choosing such a coordinate system is that the brane moves during
the cosmological evolution and the junction conditions are difficult
to implement.

In two--brane systems, the situation is not easier: whatever coordinates
one chooses, there are two junction conditions to be fulfilled.

We discuss now the brane point of view in more detail.

\subsubsection{The brane point of view}
We have seen that Einstein's equations on the brane have the form
\begin{equation}
G_{\mu\nu} = 8\pi G \tau_{\mu\nu} - \Lambda_{4} h_{\mu\nu} +
\kappa_5^4 \pi_{\mu\nu} - E_{\mu\nu}
\end{equation}
The projected Weyl tensor is tracefree: $E^{\mu}_{\mu} = 0$. Let us therefore
take the point of view that $E_{\mu\nu}$ represents a new energy--momentum 
tensor, representing the influence of the bulk gravitational field. The fluid 
corresponding to $E_{\mu\nu}$ has been named ``Weyl--fluid''. The most 
significant contribution from this fluid is its anisotropic stress 
$\delta \pi_{\rm Weyl}$. Consider the off--diagonal
component of the perturbed Einstein equation. Choosing the longitudinal gauge, 
we can write the perturbed metric on the brane as
\begin{equation}
ds^2 = -(1-2\psi)dt^2 + (1+2\phi)dx^2.
\end{equation}
Assuming that the anisotropic stress of the normal matter is zero, the off--diagonal component
of Einstein's equation leads to \cite{rsperturbations}
\begin{equation}
\phi + \psi = - 8\pi G \delta \pi_{\rm Weyl}.
\end{equation}
Therefore, even in the absence of anisotropic stress of matter, the Weyl 
fluid induces an anisotropic stress, so that $\phi + \psi \neq 0$. In 
addition, there is an entropy perturbation
induced by the Weyl fluid. These are the two effects the bulk gravitational
field has on cosmological perturbations. In particular, it might leave a 
measurable signal in the microwave background radiation.

The Sachs--Wolfe effect, relating temperature fluctuations at the last 
scattering surface to metric/matter perturbations, reads now in the brane 
world theory we consider 
\cite{rsperturbations}
\begin{equation}\label{aniso}
\frac{\delta T}{T} = \left(\frac{\delta T}{T}\right)_{\rm GR}
- \frac{8\rho_{\rm rad}}{3\rho_{\rm cdm}}{\cal S^*} 
- 8\pi G \delta \pi_{\rm Weyl}
+ \frac{16\pi G}{a^{5/2}}\int da{\mbox{ }} a^{5/2} \delta \pi_{\rm Weyl}
\end{equation}
The first term is the usual expression for the temperature fluctuation 
obtained in General Relativity.
The second term includes an entropy perturbation ${\cal S^*}$, 
induced by the Weyl fluid and is related to $\delta \rho_{\rm Weyl}$ and 
$\rho =\rho_{\rm rad} + \rho_{\rm cdm}$ by
${\cal S^*} = \delta \rho_{\rm Weyl}/\rho_{\rm Weyl} - 
\delta \rho/(3(\rho + p))$. The last term, however, 
represents an integral over the history of the anisotropic stress of 
the Weyl fluid. Because there is no equation for $\pi_{\rm Weyl}$ when
considering the brane alone, one cannot make a prediction for the temperature 
anisotropy of the cosmic microwave background radiation. For this, the full 
five--dimensional problem has to be solved.

\subsection{Scalar Perturbations: CMB Anisotropies}

However, some interesting progress has been made by using low--energy 
approximations discussed in the last section. As we have seen, one considers 
scales much below the brane tension and scales much larger than the AdS 
curvature scale. The equation for the Weyl tensor reads then
\begin{equation}\label{weyl}
\nabla^\mu E_{\mu\nu} = 0.
\end{equation}
The idea is to introduce a second brane
(so--called regulator brane) in the bulk, on which proper boundary conditions 
can be chosen. The Weyl tensor is then specified by the interbrane distance 
and the energy--momentum tensor of matter on both branes. The evolution 
equation for the interbrane distance is then obtained from eq. (\ref{weyl}).

To be more concrete, consider again the five--dimensional line element
\begin{equation}
ds^2 = e^{2\phi} dy^2 + g_{\mu\nu} (y,x^\mu) dx^\mu dx^\nu.
\end{equation}
The physical distance $D$ between the branes is given by $D=\exp (\phi) \Delta y$,
where $\Delta y$ is the coordinate distance. Remember that
we have defined $\Psi = 1 - \exp(-2 e^\phi)$. Hence, $\Psi$ is completely 
specified by the interbrane distance. As established in the last section on 
the effective action, both the projective approach and the moduli space 
approximation lead to the conclusion
that the radion couples to the trace of the energy--momentum tensor and 
therefore that the theory has much in common with a scalar--tensor theory 
(see in particular eq. (\ref{weylproj})).
Thus, fluctuations of the radion satisfy a wave equation which is similar to 
that of the scalar field in a scalar--tensor theory. This then 
specifies the full evolution of the Weyl tensor and, in particular, one is 
able to calculate the effective anisotropic stress $\pi_{\rm \delta Weyl}$ 
induced by the bulk gravitational field, so that one can calculate the 
temperature fluctuations (see eq. (\ref{aniso})) completely.

So far, two cases have been studied in the literature. Let us
describe them briefly.

In the first case, the second brane was regarded as a regulator brane in order to deal with
boundary conditions in the bulk \cite{koyamacmb}. The interbrane distance was kept fixed, which can be
obtained by choosing $\rho_{reg.brane} = - \rho_{our.brane}$. Then, assuming adiabatic
perturbations, there is only one remaining free parameter specifying the amplitude of
fluctuations in the dark radiation: $\delta C_{\rm Weyl} = \delta \rho_{\rm Weyl}/\rho_{\rm rad}$.
The results are shown in the figure 3.

\begin{figure}[!ht]
\centering
\includegraphics[width=10cm,height=7cm]{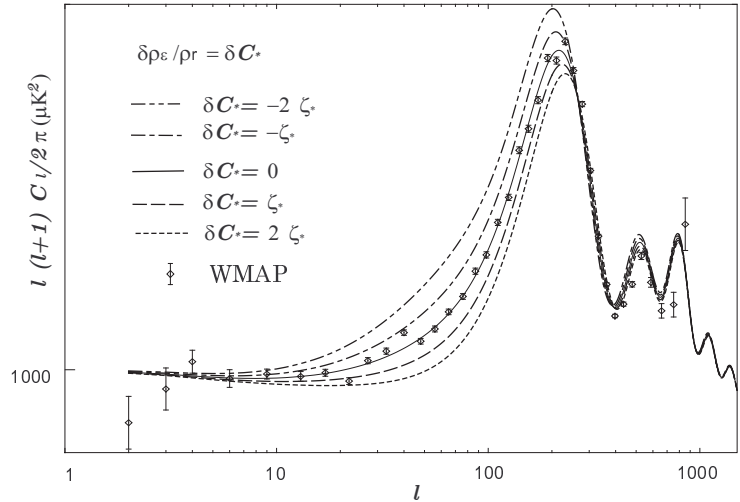}
\caption{Effects of the bulk gravitational field in brane worlds for
constant radion and adiabatic perturbations. The quantity $\delta C_{\rm Weyl}$
quantifies the amplitude of the perturbations in the Weyl fluid. $\zeta_*$ denotes here
the curvature perturbation due to dark radiation. Taken from \cite{koyamacmb}.}
\end{figure}

It was found in this case that perturbations on large angular scales
are affected most (i.e. multipole numbers smaller than 200), whereas
the effect of the Weyl fluid is small on smaller scales.

Let us now discuss the second case studied in the literature \cite{chriscmb}. Here,
the branes are free to move and there is no potential energy for the radion. The negative 
tension brane contains no matter. The background evolution
was studied in the last section. Hence, this setup reflects a two--brane system with specific matter 
content on the two branes. For the perturbations, it was assumed that 
initially, fluctuations in the radion do not contribute to the total 
curvature perturbation and that they are subsequently sourced by fluctuations 
in matter. The other perturbations between matter and radiation are assumed 
to be adiabatic. The results are shown in the figure 4.

\begin{figure}
\centering
\includegraphics[width=7cm,angle=270]{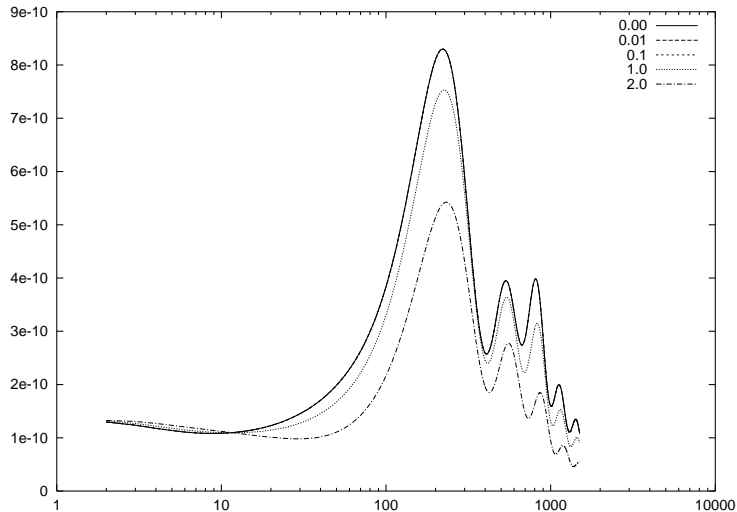}
\caption{Influence of the radion on the CMB anisotropies in a two brane system in
which the radion is allowed to move. The values in the legend denote the initial
value of the radion in Planck units. During cosmological evolution, the radion is
driven towards zero. The spectra are normalised to COBE. Taken from \cite{chriscmb}.}
\end{figure}

The case of $R_{ini}=0$ corresponds to the $\Lambda$CDM case. It can be seen from the
normalised spectra that all scales are influenced and that the spectra for
$R_{ini} >0 $ are damped, compared to the $\Lambda$CDM case.

It is not surprising that the resulting
CMB power spectra look different for the two cases taken from the literature. 
In both cases, a different behaviour of
the radion is assumed and also the nature of fluctuations in both cases is 
different.
Thus, the behaviour of the projected Weyl tensor in both cases is different.

To summarise, although some progress has been made, it is clear that brane 
worlds will not make a definite prediction for CMB perturbations. A lot will 
depend on the dynamics of the two branes, if the radion has a potential or 
not, if the perturbations in the radion initially play a role and whether or
not there is matter on the second brane. However, we will be able 
to constrain different models for the initial conditions in the very early 
universe and, maybe, different models for inflation and the cyclic universe.

\subsection{Tensor Modes}
We have seen in section 2 that, during inflation, heavy modes decay and
the zero mode is the only one which survives. However, are heavy modes
generated during the radiation or matter dominated
epochs? As is the case in General Relativity, the zero mode remains frozen outside the
horizon, but new effects might appear when the massive modes re-enter the horizon.
The only way to study this effect is to study the five--dimensional problem. To
lowest order, one can employ a ``near--brane and late--time approximation'' and
study the solution of the bulk equation near the brane \cite{tensor1},\cite{tensor2}. 
In this case, the
bulk equation is separable and can be solved explicitly. It was shown that the massive
modes decay even outside the horizon \cite{tensor2}. However, nothing can be said about 
the creation of heavy modes, since this is a higher--order effect. The problem was 
attacked analytically in \cite{tensor3} and numerically in \cite{tensor4}. 
It was shown that massive modes are sourced 
by the initial zero mode, which re-enters the horizon and is subsequently
damped. This effect has been studied numerically in \cite{tensor4} 
and the result is shown in figure $5$.

\begin{figure}[!ht]
\centering
\includegraphics[width=13cm,height=11cm]{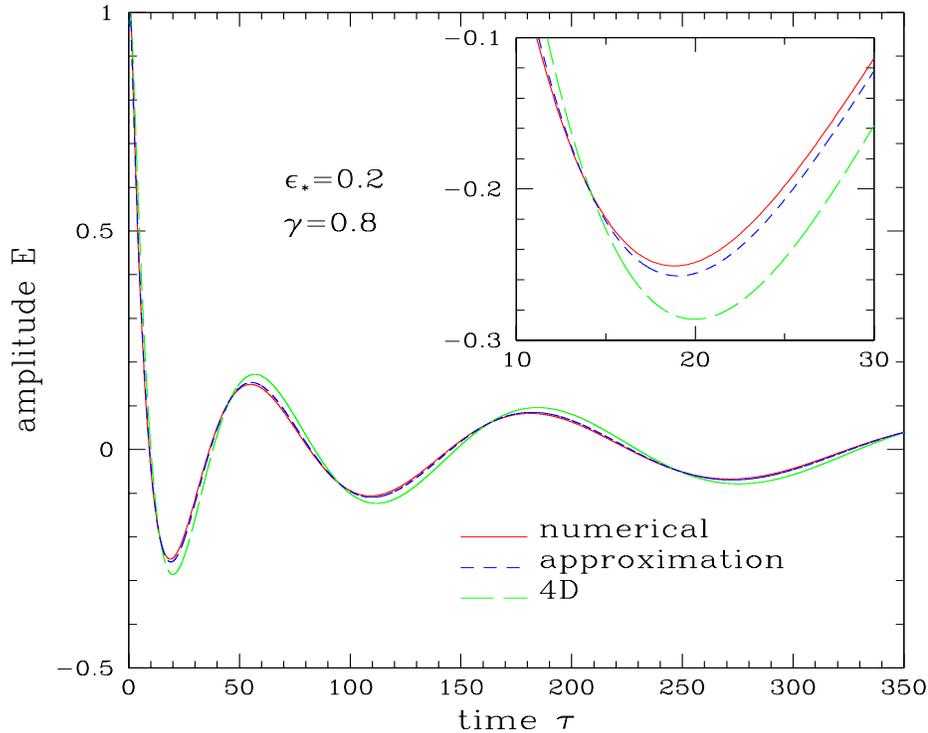}
\caption{Damping of tensor modes in the Randall--Sundrum brane world due to the
creation of massive modes. Here, $\epsilon_* = \rho_0/\lambda$ at the start of
integration and $\gamma$ is a parameter specifying the brane position. Taken 
from \cite{tensor4}.}
\end{figure}

\section{Brane Collisions}
The brane dynamics may lead to a collision between branes, either boundary
branes in the Randall-Sundrum
model or a floating five--brane with one of  the two boundary branes
for the heterotic M-theory. This possibility has been in particular 
investigated in the context of early universe cosmology 
\cite{collisionbegin}-\cite{collisionend}.
In all these cases, the brane collision is not well described at the
level of a five--dimensional effective field
theory. In particular, one may envisage that the collision may be modified
by the exchange of open strings or open membranes
when the branes get closer than a distance of the order of the string scale.
In that case, one may have a correct description of the collision. Several
scenarios have been considered.
One appealing possibility is that the singular behaviour at the brane
collision is resolved and a finite bounce happens, i.e.
the two branes approach within a minimal distance and rebound before
flying away in opposite directions.
Another possibility, which has been considered in heterotic M-theory
under the name of a ``small instanton'',
is that the impinging brane remains stuck to the other brane, implying
a change in the matter content.
Finally the two branes may go past each other, exchanging some energy
in the process. The resulting configuration
going from a big crunch to a big bang situation. This is the scenario
used in the cyclic model. 

We will first discuss the cyclic model, a brane--inspired four dimensional
model advocated to be an alternative to inflation.
These models are controversial and subject to fierce debates. In order to
illustrate the possible problems occurring during a brane collision, we
will consider the born again scenario. Again, we will be brief and 
refer to the reviews \cite{collisionreview1}-\cite{collisionreview3} for more details. 

The subject of brane collisions is still in its infancy, deserving further
study in view of its cosmological relevance. In particular, the question 
about how perturbations evolve during the brane collision is an important 
one and has sparked a lot of activity \cite{bcpertfirst}-\cite{bcpertlast}.

\subsection{The cyclic model}

The cyclic model \cite{cyclic} is an off--shoot of the ekpyrotic model 
\cite{collisionbegin}-\cite{ekpyrotic3} whereby the
big bang originated from the collision of
two nearly BPS branes. The primordial fluctuations have been argued to
follow from the small ripples on the colliding
branes. This scenario has led to heated exchanges within the cosmology community.
The cyclic scenario picks up most of the ingredients of the ekpyrotic
universe within a purely four--dimensional
description. At its most basic level, it is described by a scalar--tensor
theory involving one scalar degree of freedom which is
inspired by the radion of the Randall--Sundrum model. The potential can be
seen in figure 6 and can be separated into four different
regions. There are four phases.
Currently the field evolves slowly along a quintessence--like potential
leading to the acceleration of the expansion of the universe.
Later the field will roll down the negative part of the  potential, where
the universe undergoes a slow contracting epoch.
Quantum fluctuations in the field $\phi$ are created during this phase.
Then the field becomes kinetically dominated
and rushes towards a singularity at infinity, where it bounces back to the
region of acceleration.
The acceleration era  is useful in diluting the entropy created in each
cycle, overcoming the problems
of the cyclic universe envisaged by Tolman in the thirties.
Now this scenario relies crucially on the behaviour of the universe at
the singular bounce.

\begin{figure}[!ht]
\centering
\includegraphics[width=14cm,height=11cm]{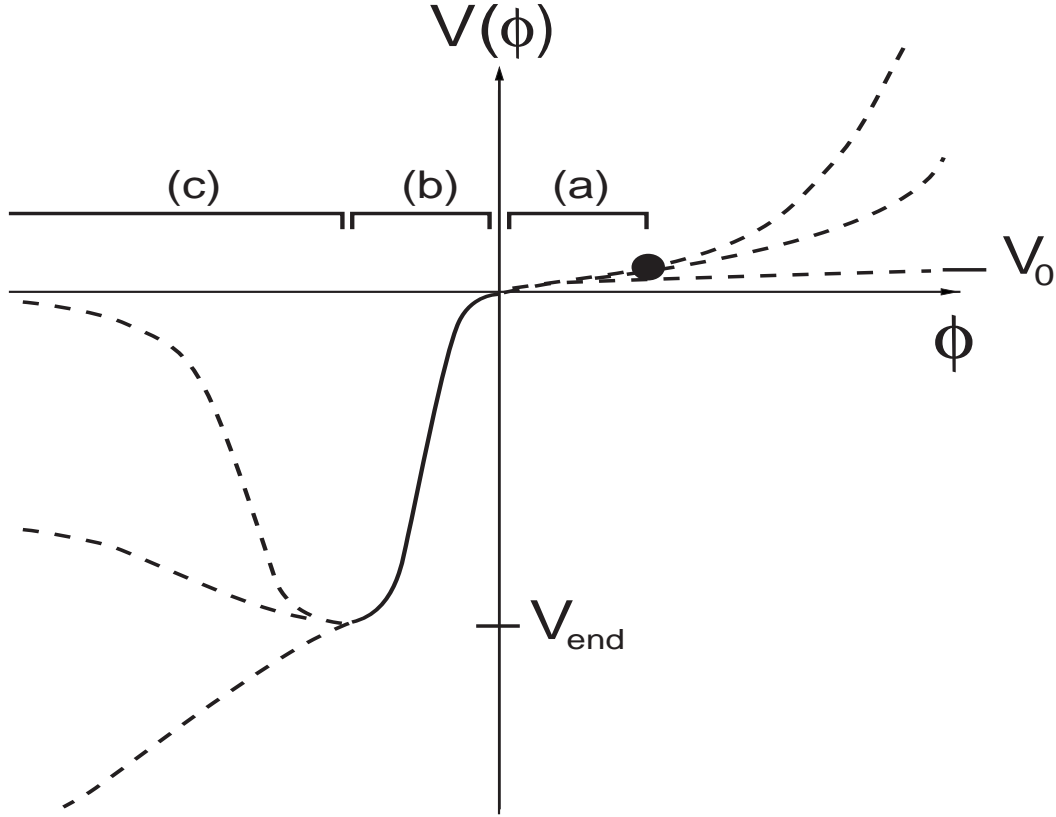}
\caption{Examples of cyclic potentials. In region (a), the potential
is such that the field leads to the present acceleration of the expansion 
of the universe. In region (b), the universe is contracting and in region (c)
the universe is driven towards the brane collision when the field value is 
$-\infty$. Taken from \cite{design}.}
\end{figure}

The bounce occurs when the field $\phi$ is infinite corresponding to the
collision of the branes.
The description of the physics at the bounce is an open question.
In the following section, we will describe an attempt to understand the
behaviour of the brane model 
at a singular bounce. It relies on the techniques, like the moduli space
approximation, that we have already presented. The model was dubbed the 
born--again brane world \cite{bornagain}.

\subsection{Born--again brane world}
The born--again brane world is a model of the early universe, in which two 
boundary branes collide. In doing so, the brane tensions change signs. 
In this model, there are no repeated cycles and as such it has to be 
distinguished from the cyclic model. 

Let us focus on the Randall--Sundrum model when the boundary branes have
detuned tensions. We have already shown that this leads to
the motion of the boundary branes when the tensions are increased compared
to the critical value corresponding to the BPS static
configuration. In particular this leads to the collision of the two boundary
branes. More precisely the Einstein frame potential reads
\begin{equation}
V(\eta)= \left( T_1  - 1 \right) \cosh^4 \left( \frac{\eta}{2}\right) 
+ \left(T_2-1\right) \sinh ^4\left( \frac{\eta}{2}\right)
\end{equation}
where $\eta$ is a normalised field related to
\begin{equation}
\eta=-\ln\vert \frac{\sqrt{1-\Psi}-1}{\sqrt{1-\Psi}+1}\vert
\end{equation}
The two branes are infinitely separated when $\eta=0$ and collide when $\eta= \infty$.
If $T_1>1$ and $T_1 +T_2 < 2$ the potential has a maximum. For $\eta$ large
enough, the branes are attracted towards each other leading to
a brane collision.
The branes collide when $\Psi =0$. Now in the brane frame of the positive
tension brane the symmetry $\Psi \to -\Psi$
exchanges positive branes and negative branes.

In a realistic setting, our universe cannot be on the negative tension brane.
Hence before the collision our brane was the negative 
tension brane which became the positive tension brane after the collision.
Before the collision the negative tension brane frame can be obtained from
the Einstein frame putting
$f_{\mu\nu}=g_{\mu\nu}/\Phi$. The field $\Phi$ vanishes linearly at the
collision leading to
a divergence of the Hubble rate in the Einstein metric. This divergence
is non--existent in the brane frame
of the negative tension brane. In that frame, the metric goes smoothly
from the pre--collision regime (a contraction in the Einstein frame) to
the post--collision regime (an expansion in the Einstein frame).
So it seems that by going to the brane frame where matter couples directly
to gravity, one can extrapolate from the pre--big bang
contracting phase to the big bang epoch.

This would justify the smooth passage used in the cyclic universe but for a 
hitch. The curvature perturbation $\zeta$ possesses a logarithmic divergence
at the collision. This divergence is present both in the
Einstein and brane frames, indicating its fundamental nature.
This indicates that the brane model is unstable at the collision. This
divergence is at the origin of the fierce debate on cosmological
perturbation in colliding brane models \cite{bcpertfirst}-\cite{bcpertlast}.

In conclusion, there is yet no known prescription to go through a cosmological
singularity such as the one springing from the collision
of two branes (for some recent progress, however, see \cite{tolley} and 
\cite{robert}. Even in the simplest setting such as the born--again brane world
scenario, divergences show an intrinsic instability
of the system. One may hope that by going to string theory  
new results on the resolution of cosmological singularities could be obtained.

\section{Summary}

In this review we have discussed aspects of brane world cosmology,
covering both the simplest Randall--Sundrum model and models with a bulk
scalar field, the latter being motivated by supersymmetry. We have seen 
that these models are formulated in such a 
way that their predictions can now be tested against precision 
data. We have discussed the early time cosmology when the Friedmann 
equation is modified and considered its effect on inflation, considered 
modifications to general relativity and how this constrains brane world 
models and discussed also the CMB predictions. More theoretical questions 
as well as speculative ideas concerning the very origin of the universe 
have also been considered. 
However, our review is certainly not complete. For example, we have been 
unable to cover some recent developments, such as brane cosmology in 
Gauss--Bonnet theory \cite{gbfirst}-\cite{gblast} and the cosmological 
consequences of induced gravity on the brane \cite{corrbegin}-\cite{corrend}. 
These topics give rise to interesting cosmological consequences; we refer 
to the original literature for more details. Similarly, several authors
have studied brane models in dimensions higher than five (see e.g. 
\cite{sixd1}-\cite{sixd2}). 
In this case the compactification is different as is the recovery
of Einstein gravity. Covering this is beyond the scope of this review.
However, we noted that in six--dimensional models with a bulk scalar field the 
problem with the self--tuning scenario is alleviated \cite{nilles2}. 

Brane world cosmology is in its infancy and there are many open questions.
Here we summarise some of them.

\begin{itemize}
\item In the case of the Randall--Sundrum model, 
the homogeneous cosmological evolution is well understood. However,
the evolution of cosmological perturbations is still not fully
understood. This is because the effects of the bulk gravitational field, 
encoded in the  projected Weyl--tensor, on CMB physics and Large Scale 
Structures are unknown. Whilst progress has been made 
in this direction, as discussed in section $6$, there is still much to do.

\item For models with bulk scalar fields, we have presented some 
results on the cosmological evolution of a homogeneous brane where we assumed 
that the bulk scalar field does not vary strongly around the brane; this
needs investigating.
Furthermore, for models with two branes, the cosmology has only been explored 
in the low energy regime, for instance with the moduli space approximation.
As yet the cosmology of the high energy regime is an open question and can
only be explored using techniques which go beyond the low energy effective
action. 

\item Both the bulk scalar field as well as the interbrane distance in 
two brane models could play an important role at least during some part 
of the cosmological history. For example one of the fields could play the 
role of dark energy. In that case, it is only natural 
that masses of particles, as well as other parameters, will vary. Preliminary
work in this direction was discussed in section $5$. However, more work
needs to be done in this regard.

\item The bulk singularity seems to play an important role in a cosmological 
setting. We have seen in section 5 that the negative tension brane moves 
towards the bulk singularity. Consequently cosmology requires a detailed
understanding of the singularity. For example, the brane might be repelled 
by the singularity; if this were the case it may have cosmological
consequences. More understanding of this is needed.

\item Brane collisions provide an exciting development into the possible
origin of the universe. However, much work is needed on how cosmological
perturbations arise in this scenario and how such perturbations evolve 
before and after the bounce \cite{bcpertfirst}-\cite{bcpertlast}. The problem
is again one of the singularity. Whilst progress has been made 
there are still unanswered questions and one may well have to go beyond the
current brane description and into the string theory to understand this 
singularity.

\item To date brane worlds are phenomenological models and have yet to 
make contact with an underlying theory. For example, whilst they are
motivated by string theory the current models have not arisen as an
approximation to string theory. Similarly, they have yet to be embedded
in a realistic theory of particle physics. Much work is needed in this
direction.
\end{itemize}

Brane cosmology has opened up new horizons, enabling theories in extra
dimensions to be explored and tested. Many interesting questions have
arisen. With new techniques and high precision data cosmology will play 
an increasingly important role in testing ideas beyond the standard
model of particle physics.

\vspace{0.5cm}

\noindent{\bf Acknowledgements:} We are grateful to our colleagues and friends 
for discussions and comments on brane cosmology. In particular we would
like to thank Sam Webster for his comments on a previous version.
This work was supported in part by PPARC.


\section*{References}
\begin{thebibliography}{300}
\bibitem{Kaluza} Kaluza, T. 1921 {\it Sitzungsber.Preuss.Akad.
Wiss.Berlin (Math.Phys.) K1} p 966; Klein O 1926 Z.Phys. {\bf 37} 895
\bibitem{polchinskibook} Polchinski J. 1999 {\it String Theory, Two
Volumes}, Cambridge University Press
\bibitem{horavawitten} Horava P, Witten E 1996 Nucl.Phys.B{\bf 460}
506; ibid Nucl.Phys.B{\bf 475} 94
\bibitem{witten} Witten E 1996 Nucl. Phys.B {\bf 471} 135
\bibitem{lukasstelle} Lukas A, Ovrut B A, Stelle K S, Waldram D 1999
Phys.Rev.D{\bf 59} 086001; ibid Nucl.Phys.B {\bf 552} 246
\bibitem{akama} Akama K 1982 Lect.Notes Phys {\bf 176} 267
\bibitem{rubakov} Rubakov V A, Shaposhnikov M E 1983 Phys.Lett.B {\bf
125} 136
\bibitem{visser} Visser M 1985 Phys.Lett.B {\bf 159} 22
\bibitem{squires} Squires E J 1986 Phys.Lett.B {\bf 167} 286
\bibitem{gibbons} Gibbons G W, Wiltshire D L 1987
Nucl. Phys. B{\bf 717}
\bibitem{arkanihamed1} Arkani-Hamed N, Dimopoulos S, Dvali G 1998
Phys.Lett.B{\bf 429} 263
\bibitem{arkanihamed2} Antoniadis I, Arkani-Hamed N, Dimopoulos S, Dvali G 1998
Phys.Lett.B{\bf 436} 257
\bibitem{antoniadis} Antoniadis I 1990 Phys.Lett.B{\bf 246} 377
\bibitem{randallsundrum1} Randall L, Sundrum R 1999
Phys.Rev.Lett. {\bf 83} 3370
\bibitem{randallsundrum2} Randall L, Sundrum R 1999
Phys.Rev.Lett. {\bf 83} 4690
\bibitem{goldbergerwise} Goldberger W D, Wise M B 1999 Phys.Rev.Lett.
{\bf 83} 4922
\bibitem{binetruy1} Binetruy P, Deffayet C, Langlois D 2000
Nucl.Phys.{\bf 565} 269
\bibitem{boundaryinflation} Lukas A, Ovrut B A, Waldram D 1999
Phys.Rev.D {\bf 61}
\bibitem{grojean} Cline J M, Grojean C, Servant G 1999
Phys.Rev.Lett. {\bf 83} 4245
\bibitem{csaki} Csaki C, Graesser M, Kolda C, Terning J 1999
Phys.Lett.B {\bf 462} 34
\bibitem{selftuning} Arkani-Hamed N, Dimopoulos S, Kaloper N, Sundrum R
2000 Phys. Lett.B {\bf 480} 193
\bibitem{nilles} Forste S, Lalak Z, Lavignac S, Nilles H-P 2000
Phys.Lett.B {\bf 481} 360
\bibitem{ourreview} Brax Ph, van de Bruck C, Class.Quant.Grav. {\bf 20}, R201 (2003)
\bibitem{luki} Lukas A, Ovrut B A, Waldram D 1998 hep-th/9812052
\bibitem{rubi} Rubakov V A 2001 Phys.Usp.{\bf 44} 871
\bibitem{damien} Easson D A 2000 Int.J.Mod.Phys. A{\bf 16} 4803
\bibitem{wandi} Wands D 2002 Class.Quant.Grav.{\bf 19} 3403
\bibitem{langi} Langlois D 2002 Prog. Theo. Phys. Suppl. {\bf 148} 181
\bibitem{padilla} Padilla A, PhD thesis 2002, University of Durham, hep-th/0210217
\bibitem{gabi} Gabadadze G, hep-ph/0308112
\bibitem{maartens} Maartens R, gr-qc/0312059
\bibitem{quevedo} Quevedo F 2002 Class.Quant.Grav. {\bf 19} 5721
\bibitem{submillimeter} Hoyle C D, et al 2001 Phys.Rev.Lett. {\bf 86}
1418
\bibitem{flanagan} Flanagan E E, Tye S H H, Wasserman I 2000
Phys.Rev.D {\bf 62} 044039
\bibitem{vandebruck1} van de Bruck C, Dorca M, Martins C, Parry M 2000
Phys.Lett.B {\bf 495} 183
\bibitem{darkradiation1} Mukohyama S 1999 Phys.Lett.B {\bf 473} 241
\bibitem{darkradiation2} Ida D 2000 JHEP {\bf 0009} 014
\bibitem{darkradiation3} Ichiki K, Yahiro M, Takino T, Orito M,
Mathews G J 2002 Phys.Rev.D {\bf 66} 043521
\bibitem{shirubulk} Mukohyama S, Shiromizu T, Maeda K-I 2000
Phys.Rev.D {\bf 62} 024028
\bibitem{kraus} Kraus P 1999 JHEP {\bf 9912} 011
\bibitem{binetruy2} Binetruy P, Deffayett C, Ellwanger U, Langlois D 2000
Phys.Lett.B {\bf 477} 285
\bibitem{charmousis} Bowcock P, Charmousis C, Gregory R 2000
Class.Quant.Grav.{\bf 17} 4745
\bibitem{maartensw} Maartens R 2001 gr-qc/0101059
\bibitem{shiromizu} Shiromizu T, Madea K, Sasaki M 2000
Phys.Rev.D {\bf 62} 024012
\bibitem{wald} Wald R 1984 {\it General Relativity} University of Chicago Press; 
Carroll S, {\it Space--time and Geometry: Introduction to General Relativity}, 
Addison Wesley (2003)
\bibitem{israel} Israel W 1966 Nuovo Cim B{\bf 44S10} 1;
Erratum: ibid Nuovo Cim B{\bf 48} 463
\bibitem{roy} Maartens R 2000 Phys.Rev.D {\bf 62} 084023
\bibitem{wandsinflation} Maartens R, Wands D, Bassett B A,
Heard I 2000 Phys.Rev.D{\bf 62} 041301
\bibitem{copeland} Copeland E J, Liddle A R, Lidsey J E 2001
Phys.Rev.D {\bf 64} 023509
\bibitem{anne} Davis S C, Perkins W B, Davis A-C, Vernon I R 2000
Phys.Rev.D{\bf 63} 083518
\bibitem{rsperturbations} Langlois D, Maartens R, Sasaki M, Wands D 2001
Phys.Rev.D{\bf 63} 084009
\bibitem{bardeen} Bardeen J M, Steinhardt P J, Turner M S
1983 Phys.Rev.D{\bf 28} 679
\bibitem{malik} Wands D, Malik K, Lyth D H, Liddle A R (2000)
Phys.Rev.D{\bf 62} 043527
\bibitem{liddlebook} Liddle A R, Lyth D 2000 {\it Cosmological
Inflation and Large Scale Structure} Cambridge University Press
\bibitem{langloisperturbations} Langlois D, Maartens R, Wands D 2000
Phys.Lett.B {\bf 489} 259
\bibitem{rubakovperturbations} Gorbunov D S, Rubakov V A,
Sibiryakov S M (2001) J. High Energy Phys. JHEP 10(2001)015
\bibitem{lidsey} Huey G, Lidsey J E 2002 Phys.Rev.D {\bf 66} 043514
\bibitem{calgagni1} Calgagni G 2003 JCAP {\bf 0311}:009 
\bibitem{calgagni2} Calgagni G 2003, hep-ph/0312246
\bibitem{calgagni3} Calgagni G 2004, hep-ph/0402126
\bibitem{seery} Seery D, Taylor A 2003, astro-ph/0309512
\bibitem{shinji} Tsujikawa S, Liddle A R (2004) JCAP {\bf 0403} 001
\bibitem{gordon} Gordon D, Wands D, Bassett B A, Maartens R (2001)
Phys.Rev.D{\bf 63} 023506
\bibitem{ashcroft1} Ashcroft P, van de Bruck C, Davis A C 2002
Phys.Rev.D {\bf 66} 121302
\bibitem{riotto} Wands D, Bartolo, N, Matarrese, S, Riotto A. 2002 
Phys.Rev.D {\bf 66} 043520
\bibitem{emparan} Emparan R, Horowitz G T, Myers R C 2000 Phys.Rev.Lett. {\bf 85} 499
\bibitem{guedens1} Guedens R, Clancy D, Liddle A R 2002
Phys.Rev.D{\bf 66} 043513; 
\bibitem{guedens2} Guedens R, Clancy D, Liddle A R 2002 
Phys.Rev.D{\bf 66} 083509
\bibitem{hawking} Carr B J, Hawking S W 1974 MNRAS {\it 168} 399 
\bibitem{guedens3} Clancy D, Guedens R, Liddle A R 2003 
Phys.Rev.D{\bf 68} 023507
\bibitem{bhfirst} Majumdar A S 2003 Phys.Rev.Lett. {\bf 90} 031303
\bibitem{kloper} Emparan R, Garcia-Bellido J, Kaloper N 2003 JHEP {\bf 0301} 079
\bibitem{bhlast} Majumdar A S, Mukherjee N 2004 astro-ph/0403405
\bibitem{bhreview} Kanti P 2004 hep-ph/0402168
\bibitem{graviton1} Langlois D, Sorbo L, Rodriguez-Martinez M 2002 
Phys.Rev.Lett.{\bf 89} 171301 
\bibitem{graviton2} Langlois D, Sorbo L 2003 Phys.Rev.D {\bf 68} 084006 
\bibitem{leeper} Leeper E, Maartens R, Sopuerta C F 2004 Class.Quant.Grav.{\bf 21} 1125
\bibitem{Ross} Kogan I I, Ross G G 2000 Phys.Lett.B {\bf 485} 255
\bibitem{tye} Stoica H, Tye S H H, Wasserman I 2000 Phys.Lett.B {\bf 482} 205
\bibitem{anne2} Davis A-C, Davis S C, Perkins W B, Vernon I R 2001 Phys.Lett.B {\bf 504}, 254
\bibitem{gubser} Gubser S S 2001 Phys.Rev.D{\bf 63} 084017
\bibitem{csakiholo} Csaki C, Erlich J, Hollowood T J, Terning J 2001
Phys.Rev.D{\bf 63} 065019
\bibitem{verlinde} Verlinde E 2000 hep-th/0008140
\bibitem{tanaka} Tanaka T 2002 gr-qc/0203082
\bibitem{kaloper1} Emparan R, Fabbri A, Kaloper N 2002 JHEP {\bf 0208} 043
\bibitem{kaloper2} Emparan R, Garcia-Bellido J, Kaloper N 2002 hep-th/0212132
\bibitem{thooft} 't Hooft G 2000 hep-th/0003004; Bousso R 2002 Rev.Mod.Phys.{\bf 74}, 825
\bibitem{hawking1} Hawking S W, Hertog T, Reall H S 2000 Phys.Rev.D{\bf 62} 043501
\bibitem{hawking2} Hawking S W, Hertog T, Reall H S 2001 Phys.Rev.D{\bf 63} 083504
\bibitem{tracefirst} Nojiri S, Odinstov S D 2000 Phys.Lett.B{\bf 484} 119
\bibitem{tracelast} Anchordoqui L, Nunez C, Olsen K 2000 JHEP {\bf 0010} 050
\bibitem{maedawands} Maeda K, Wands D 2000 Phys.Rev.D{\bf 62} 124009
\bibitem{mennim} Mennim A, Battye R A 2001 Class.Quant.Grav.{\bf 18} 2171
\bibitem{kanti1} Kanti P, Olive K A, Pospelov M 2000
Phys.Lett.B{\bf 481} 386
\bibitem{langloisbulkscalar} Langlois D, Rodriguez-Martinez M 2001
Phys.Rev. D{\bf 64} 123507
\bibitem{flanagan2} Flanagan E E, Tye S H H, Wasserman I 2001
Phys.Lett.B{\bf 522} 155
\bibitem{chamblin} Chamblin H A, Reall H S 1999 Nucl.Phys. {\bf 562} 133
\bibitem{kunze} Kunze K E, Vazquez-Mozo M A 2002 Phys.Rev.D{\bf 65} 044002
\bibitem{stephen1} Davis S C 2002 JHEP {\bf 0203} 054
\bibitem{stephen2} Davis S C 2002 JHEP {\bf 0203} 058
\bibitem{kanti2} Kanti P, Lee S, Olive K A 2002 hep-th/0209036
\bibitem{karch} DeWolfe O, Freedman D Z, Gubser S S, Karch A 2000
Phys.Rev.D {\bf 62} 046008
\bibitem{brax1} Brax Ph, Davis A C 2001 Phys.Lett.B {\bf 497}
289
\bibitem{groje} Csaki C, Erlich J, Grojean C, Hollowood T J 2000
Nucl. Phys. B{\bf 584} 359
\bibitem{youm} Youm D 2001 Nucl. Phys. B{\bf 596} 289
\bibitem{braxsingularity} Brax Ph, Davis A C 2001 Phys.Lett.B
{\bf 513} 156
\bibitem{br1} Ph. Brax, A. Falkowski and Z. Lalak, Phys. Lett. {\bf B521} (2001) 105.
\bibitem{br2} Ph. Brax and N. Chatillon, JHEP 0312 (2003) 026.
\bibitem{bscosfirst} Mohapatra R N, Perez-Lorenzana A,
de Sousa Pires C A 2000 Phys.Rev.D{\bf 62} 105030
\bibitem{himo} Himemoto Y, Sasaki M 2002
Phys.Rev.D {\bf 63} 044015
\bibitem{hime} Himemoto Y, Tanaka T, Sasaki M 2002
Phys.Rev.D {\bf 65} 104020
\bibitem{koyama2} Koyama K, Takahashi K 2003 hep-th/0301165
\bibitem{bscoslast} Wang B, Xue L, Zhang X, Hwang W YP 2003
hep-th/0301072
\bibitem{brax2} Brax Ph, Davis A C 2001 J. High Energy Phys.
JHEP 05(2001)007
\bibitem{vandebruck2} Brax Ph, van de Bruck C, Davis A C 2001
J. High Energy Phys. JHEP 10(2001)026
\bibitem{chiba} Chiba, T 2001 gr-qc/0110118
\bibitem{uzan} Uzan J-P 2003 Rev.Mod.Phys. {\bf 75}, 403
\bibitem{fujiibook} Fujii Y, Maeda K-I 2003 {\it The Scalar-Tensxor Theory
of Gravitation} Cambridge University Press
\bibitem{kiritsis} Kiritsis E, Kofinas G, Tetradis N, Tomaras T N, Zarikas V 2002
JHEP {\bf 0302} 035
\bibitem{tetradis} Tetradis N 2002 Phys.Lett.B{\bf 569} 1
\bibitem{braxlanglois} Brax Ph, Langlois D, Rodriguez-Martinez M 2002
Phys.Rev.D{\bf 67} 104022
\bibitem{garriga} Garriga J, Tanaka T 2000 Phys.Rev.Lett.{\bf 84} 2778
\bibitem{chiba2} Chiba T 2000 Phys.Rev.D{\bf 62} 021502
\bibitem{brax3} Brax Ph, van de Bruck C, Davis A C,
Rhodes C.S. 2002 Phys.Lett.B {\bf 531} 135
\bibitem{brax4} Brax Ph, van de Bruck C, Davis A C,
Rhodes C.S. 2002 Phys.Rev.D {\bf 65} 121501
\bibitem{vandebruck3} Brax Ph, van de Bruck C, Davis A C,
Rhodes C.S. 2003 Phys.Rev.D{\bf 67} 023512
\bibitem{cynolter} Cynolter G 2002 hep-th/0209152
\bibitem{radcosfirst} Csaki C, Graesser M l, Randall L, Terning J 1999
Phys.Rev.D {\bf 62} 045015
\bibitem{clinerad} Cline J, Firouzjahi H 2001 Phys.Rev.D{\bf 64} 023505
\bibitem{clinerad2} Cline J, Firouzjahi H 2000 Phys.Lett.B {\bf 495}, 271
\bibitem{radioncharmousis} Charmousis C, Gregory R, Rubakov V A 2000
Phys.Rev.D{\bf 62} 067505
\bibitem{greasser} Csaki C, Graesser M L, Kribs G D 2001
Phys.Rev.D{\bf 63} 065002
\bibitem{gen} Gen U, Sasaki M 2001 Prog.Theor.Phys. {\bf 105} 591
\bibitem{nolte} Grinstein B, Nolte D R, Skiba W 2001
Phys.Rev.D{\bf 63} 105016
\bibitem{langrad} Binetruy P, Deffayet C, Langlois D 2001
Nucl.Phys.B{\bf 615} 219
\bibitem{anupam} Mazumdar A, Perez-Lorenzana A 2001
Phys.Lett.B{\bf 508} 340
\bibitem{anupam2} Mazumdar A, Mohapatra R N, Perez-Lorenzana A 2003 
hep-ph/0310258
\bibitem{verni} Vernon I R, Davis A.-C. 2004 hep-ph/0401201
\bibitem{sorboz} Langlois D, Sorbo L 2002 Phys.Lett.B{\bf 543} 155
\bibitem{kofi} Kofman L, Martin J, Peloso M 2004 hep-ph/0401189
\bibitem{radcoslast} Lesgourgues J, Sorbo L 2003 hep-th/0310007
\bibitem{barrow} Webb J K, Murphy M T, Flambaum V V, Dzuba V A,
Barrow J D, Churchill C W, Prochaska J X, Wolfe A M 2001
Phys.Rev.Lett. {\bf 87} 091301
\bibitem{vinet} Cline J M, Vinet J 2002 JHEP {\bf 0202}, 042
\bibitem{braneaction0} Khoury J, Zhang R 2002
Phys.Rev.Lett.{\bf 89} 061302
\bibitem{braneaction1} Kanno S, Soda J 2002
Phys.Rev.D{\bf 66} 083506
\bibitem{braneaction2} Kobayashi S, Koyama K 2002 JHEP {\bf 0212} 056
\bibitem{braneaction3} Shiromizu T, Koyama K 2002 Phys.Rev.D{\bf 67} 084022
\bibitem{braneaction3a} Shiromizu T, Koyama K 2003 Phys.Rev.D{\bf 67} 104011
\bibitem{braneaction4} Wiseman T 2002 Class.Quant.Grav. {\bf 19}
3083
\bibitem{gilles} Damour T, Esposito-Farese G 1992
Class.Quant.Grav.{\bf 9} 2093
\bibitem{damour} Damour T, Nordtvedt K 1993 Phys.Rev.Lett.{\bf 70} 2217
\bibitem{ashcroft2} Ashcroft P R, van de Bruck C, Davis A.-C. 2003 
astro-ph/0310643, to appear in Phys.Rev.D
\bibitem{varyour} Brax Ph, van de Bruck C, Davis A-C, Rhodes C S 2003
Astrophys.Space Sci. {\bf 283} 627
\bibitem{palma} Palma G A, Brax Ph, Davis A-C, van de Bruck C 2003 Phys.Rev.D{\bf 68} 123519
\bibitem{kaluza-klein} Kanno S, Soda J 2003 hep-th/0312106
\bibitem{proapproachini} Kanno S, Soda J 2004 Gen.Rel.Grav.{\bf 36} 689
\bibitem{dbrane} Shiromizu T, Koyama K, Onda S, Torii T 2003 
Phys.Rev.D {\bf 68} 063506
\bibitem{dbrane1a} Shiromizu T, Koyama K, Torii T 2003 Phys.Rev.D{\bf 68} 103513
\bibitem{dbrane2} Onda S, Shiromizu T, Koyama K, Hayakawa S 2003 hep-th/0311262
\bibitem{gravifield} Yoshiguchi H, Koyama K 2004 hep-th/0403097
\bibitem{proapproachfini} Kanno S, Soda J 2003 hep-th/0312106
\bibitem{perturbationsfirst} Mukohymama S 2000 Phys.Rev.D {\bf 62} 084015
\bibitem{mukija} Mukohymama S 2001 Phys.Rev.D {\bf 64} 064006
\bibitem{mukija2}  Mukohymama S 2000 Class.Quant.Grav. {\bf 17} 4777
\bibitem{kodama} Kodama H, Ishibashi A, Seto O 2000 Phys.Rev.D {\bf 62} 064022
\bibitem{roypertur} Maartens R 2000 Phys.Rev.D {\bf 62} 084023
\bibitem{langlispertur} Langlois D 2000 Phys.Rev.D {\bf 62} 126012
\bibitem{vdbruckpertur} van de Bruck C, Dorca M, Brandenberger R H, Lukas A 2000
Phys.Rev.D {\bf 62} 123515
\bibitem{koyama} Koyama K, Soda J 2000 Phys.Rev.D {\bf 62} 123502
\bibitem{royandchris} Gordon C, Maartens R 2001 Phys.Rev.D{\bf 63} 044022
\bibitem{langloisnew} Langlois D 2001 Phys.Rev.Lett. {\bf 86} 2212
\bibitem{sodakoyama} Kobayashi S, Koyama K, Soda J 2001 Phys.Lett.B{\bf 501} 157
\bibitem{malik2} Bridgman H A, Malik K A, Wands D 2002 Phys.Rev.D {\bf 65} 043502
\bibitem{nathalie} Deruelle N, Dolezel T, Katz J 2001 Phys.Rev.D {\bf 63} 083513
\bibitem{vdb2} van de Bruck C, Dorca M 2000 hep-th/0012073
\bibitem{vdb3} Dorca M, van de Bruck C 2001 Nucl.Phys.B {\bf 605} 215
\bibitem{ivo} Neronov A, Sachs I 2001 Phys.Lett.B {\bf 513} 173
\bibitem{malik3} Bridgman H A, Malik K A, Wands D 2001 Phys.Rev.D {\bf 63} 084012
\bibitem{radionpertur} Koyama K 2002 Phys.Rev.D{\bf 66} 084003 
\bibitem{seti} Seto O, Kodama H 2001 Phys.Rev.D {\bf 63} 123506
\bibitem{soda2} Koyama K, Soda J 2002 Phys.Rev.D{\bf 65} 023514
\bibitem{sago} Sago N, Himemoto Y, Sasaki M 2002 Phys.Rev.D{\bf 65} 024014
\bibitem{bernhard1} Leong B, Dunsby P, Challinor A, Lasenby A 2002
Phys.Rev.D{\bf 65} 104012
\bibitem{bernhard2} Leong B, Challinor A, Maartens R, Lasenby A 2002
Phys.Rev.D{\bf 66} 104010
\bibitem{deffayet} Deffayet C 2002 Phys.Rev.D {\bf 66} 103504
\bibitem{riotti} Giudice G F, Kolb E W, Lesgourges J, Riotto A 2002
Phys.Rev.D {\bf 66} 083512
\bibitem{riazbrane} Riazuelo A, Vernizzi F, Steer D, Durrer R 2002
hep-th/0205220
\bibitem{gum} Gumjudpai B, Maartens R, Gordon C 2003 Class.Quant.Grav.{\bf 20} 3295
\bibitem{perturbationslast} Ringeval C, Boehm T, Durrer R 2003 hep-th/0307100
\bibitem{nathaliereview} Deruelle N 2003 gr-qc/0301035
\bibitem{royreview} Maartens R 2004 astro-ph/0402485
\bibitem{koyamacmb} Koyama K 2003 Phys.Rev.Lett. {\bf 91} 221301
\bibitem{chriscmb} Rhodes C S, van de Bruck C, Brax Ph, Davis A.-C. 
2003 Phys.Rev.D{\bf 68} 083511
\bibitem{tensor1} Mennim, A 2003 PhD thesis, University of Cambridge;
Mennim A, Battye R A, van de Bruck C 2003 Astrophys.Space Sci.
{\bf 283} 633  
\bibitem{tensor2} Battye R A, van de Bruck C, Mennim A 2003 hep-th/0308134 to appear in 
Phys.Rev.D
\bibitem{tensor3} Easther R, Langlois D, Maartens R, Wands D 2003 JCAP {\bf 0310} 014
\bibitem{tensor4} Hiramatsu T, Koyama K, Taruya A 2004 Phys.Lett.B{\bf 578} 269 
\bibitem{collisionbegin} Khoury J, Ovrut B A, Steinhardt P J, Turok N 2001
Phys.Rev.D{\bf 64} 123522
\bibitem{ekpyrotic2} Khoury J, Ovrut B A, Seiberg N, Steinhardt P J, Turok N 2002
Phys.Rev.D{\bf 65} 086007
\bibitem{ekpyrotic3} Khoury J, Ovrut B A, Steinhardt P J, Turok N 2002
Phys.Rev.D{\bf 66} 046005
\bibitem{ekpyrotic4} Enqvist K, Keski-Vakkuri E, Rasanen S 2001
Phys.Rev.D {\bf 614} 388
\bibitem{ekpyrotic5} Kallosh R, Kofman L, Linde A D 2001
Phys.Rev.D {\bf 64} 123523
\bibitem{nero} Neronov A 2001 JHEP {\bf 0111} 007
\bibitem{rasanen} Rasanen S 2002 Nucl.Phys. B{\bf 626} 183
\bibitem{cyclic} Steinhardt P J, Turok N 2002
Science {\bf 296} 1436; ibid 2002 Phys.Rev.D {\bf 65} 126003
\bibitem{bornagain} Kanno S, Sasaki M, Soda J 2003 Prog.Theor.Phys.{\bf 109}, 357
\bibitem{transitionov} Ovrut B A, Pantev T, Park J 2000 JHEP {\bf 0005} 045
\bibitem{transitionwi} Witten E 1996 Nucl. Phys. B{\bf 460} 541
\bibitem{desaster} Horowitz G T, Polchinski J 2002 Phys.Rev.D{\bf 66} 103512
\bibitem{copebrane} Copeland E J, Gray J, Lukas A, Skinner D 2002 Phys.Rev.D{\bf 66} 124007
\bibitem{finellif} Finelli F 2003 JCAP {\bf 0310} 011
\bibitem{gray} Gray J, Lukas A, Probert G I 2003 hep-th/0312111
\bibitem{collisionend} Antunes N, Copeland E J, Hindmarsh M, Lukas A 2003 hep-th/0310103
\bibitem{collisionreview1} Rasanen S 2002 astro-ph/0208282
\bibitem{collisionreview2} Khoury J 2004 astro-ph/0401579
\bibitem{collisionreview3} Turok N, Steinhardt P J 2004 hep-th/0403020
\bibitem{bcpertfirst} Durrer R, Vernizzi F 2002 Phys.Rev.D {\bf 66} 083503
\bibitem{comment} Martin J, Peter P, Pinto-Neto N, Schwarz D J 2003 
Phys.Rev.D{\bf 67} 028301
\bibitem{note} Durrer R, hep-th/0112026
\bibitem{peter} Peter P, Pinto-Neto N 2002 Phys.Rev.D {\bf 66} 063509
\bibitem{brand} Brandenberger R, Finelli F 2001 JHEP{\bf 0111} 056
\bibitem{lyth} Lyth D H 2002 Phys.Lett.B{\bf 524} 1
\bibitem{hwang} Hwang J-C 2002 Phys.Rev.D {\bf 65} 063514
\bibitem{branden} Tsujikawa S, Brandenberger R H, Finelli F 2002
Phys.Rev.D {\bf 66} 083513 
\bibitem{gortur} Gordon C, Turok N 2003 Phys.Rev.D{\bf 67} 123513
\bibitem{parame} Martin J, Peter P, 2003 Phys.Rev.D{\bf 68} 103517
\bibitem{cartier} Cartier C, Durrer R, Copeland E J 2003 Phys.Rev.D {\bf 67} 103517
\bibitem{turi} Boyle L A, Steinhardt P J, Turok N 2003 hep-th/0307170
\bibitem{tolley} Tolley A J, Turok N, Steinhardt P J 2003 hep-th/0306109
\bibitem{craps} Craps B, Ovrut B A 2004 Phys.Rev.D{\bf 69} 066001
\bibitem{martin} Martin J, Peter P 2004 Phys.Rev.Lett.{\bf 92} 061301
\bibitem{robert} Battefeld T J, Patil S P, Brandenberger R H 2004 hep-th/0401010
\bibitem{boyle} Boyle L A, Steinhardt P J, Turok N, hep-th/0403026
\bibitem{design} Khoury J, Steinhardt P J, Turok N 2004 Phys.Rev.Lett.{\bf 92} 031302
\bibitem{bcpertlast} Martin J, Peter P 2004 hep-th/0403173
\bibitem{gbfirst} Binetruy P, Charmousis C, Davis S C, Dufaux J-F 2002
Phys.Lett.B {\bf 544} 183
\bibitem{germani} Germani C, Sopuerta C F 2002 Phys.Rev.Lett. {\bf 88} 231101
\bibitem{gbchristos} Charmousis C, Dufaux J-F 2002 Class.Quant.Grav.{\bf 19} 4671
\bibitem{gblast} Gravanis E, Willison S 2002 hep-th/0209076
\bibitem{corrbegin} Dvali G, Gabadadze G, Porrati M 2000 Phys.Lett.B{\bf 485} 208
\bibitem{hol} Collins H, Holdom B 2000 Phys.Rev.D{\bf 62} 105009
\bibitem{shta} Shtanov Yu V 2000 hep-th/0005193
\bibitem{defa} Deffayet C, Dvali G, Gabadadze G 2002 Phys.Rev.D{\bf 65} 044023
\bibitem{dia} Diamandis, Georgalas, Mavromatos N, Papantonopoulos E, Pappa I 2002 
Int.J.Mod.Phys. {\bf A 17} 2241
\bibitem{shani} Sahni V, Shtanov Yu 2002 astro-ph/0202346
\bibitem{mad} Maeda K I, Mizuno S, Torii T 2003 Phys.Rev.D{\bf 68} 024033
\bibitem{multi} Multam\"aki, Gaztanaga E, Manera M 2003 MRNAS {\bf 344} 761
\bibitem{corrend} Lue A, Starkman G 2003 Phys.Rev.D{\bf 67} 064002
\bibitem{sixd1} Aghababaie Y et al 2003 JHEP {\bf 0309} 037
\bibitem{sixa4} Schwindt, J.-M. Wetterich C 2004 Phys.Lett.B {\bf 578} 409
\bibitem{sixa3} Graesser M L, Kile J E, Wang P 2004 hep-th/0403074
\bibitem{sixa2} Navarro I, Santiago J 2004 hep-th/0402204 
\bibitem{sixa1} Lee H M, Tasinato G 2004 hep-th/0401221
\bibitem{sixd2} Battye R A, Carter B, Mennim A 2003 hep-th/0312198
\bibitem{nilles2} Nilles H-P, Papazoglou A, Tasinato G 2004 
Nucl.Phys.B{\bf677} 405
\end{thebibliography}
\end{document}